%% file: main.tex
\PassOptionsToPackage{unicode}{hyperref}
\PassOptionsToPackage{hyphens}{url}
\PassOptionsToPackage{dvipsnames,svgnames,x11names}{xcolor}
\documentclass[
  american,
  11pt,
]{article}
\usepackage{xcolor}
\usepackage[margin=0.85in,letterpaper]{geometry}
\usepackage{amsmath,amssymb}
\usepackage{iftex}
\ifPDFTeX
  \usepackage[T1]{fontenc}
  \usepackage[utf8]{inputenc}
  \usepackage{textcomp} 
\else 
  \usepackage{unicode-math} 
  \defaultfontfeatures{Scale=MatchLowercase}
  \defaultfontfeatures[\rmfamily]{Ligatures=TeX,Scale=1}
\fi
\usepackage{lmodern}
\ifPDFTeX\else
\fi
\IfFileExists{upquote.sty}{\usepackage{upquote}}{}
\IfFileExists{microtype.sty}{
  \usepackage[]{microtype}
  \UseMicrotypeSet[protrusion]{basicmath} 
}{}
\makeatletter
\@ifundefined{KOMAClassName}{
  \IfFileExists{parskip.sty}{%
    \usepackage{parskip}
  }{
    \setlength{\parindent}{0pt}
    \setlength{\parskip}{6pt plus 2pt minus 1pt}}
}{
  \KOMAoptions{parskip=half}}
\makeatother
\usepackage{color}
\usepackage{fvextra}

\DefineVerbatimEnvironment{Highlighting}{Verbatim}{commandchars=\\\{\},breaklines=true,fontsize=\small}
\newenvironment{Shaded}{}{}

\newcommand{\AttributeTok}[1]{\textcolor[rgb]{0.49,0.56,0.16}{#1}}

\newcommand{\CommentTok}[1]{\textcolor[rgb]{0.38,0.63,0.69}{\textit{#1}}}

\newcommand{\ControlFlowTok}[1]{\textcolor[rgb]{0.00,0.44,0.13}{\textbf{#1}}}

\newcommand{\FunctionTok}[1]{\textcolor[rgb]{0.02,0.16,0.49}{#1}}

\newcommand{\NormalTok}[1]{#1}
\newcommand{\OperatorTok}[1]{\textcolor[rgb]{0.40,0.40,0.40}{#1}}

\newcommand{\StringTok}[1]{\textcolor[rgb]{0.25,0.44,0.63}{#1}}

\usepackage{longtable,booktabs,array}
\usepackage{calc} 
\usepackage{etoolbox}
\makeatletter
\patchcmd\longtable{\par}{\if@noskipsec\mbox{}\fi\par}{}{}
\makeatother
\IfFileExists{footnotehyper.sty}{\usepackage{footnotehyper}}{\usepackage{footnote}}
\makesavenoteenv{longtable}
\usepackage{graphicx}
\usepackage{adjustbox}
\makeatletter
\newsavebox\pandoc@box
\newcommand*\pandocbounded[1]{
  \sbox\pandoc@box{#1}%
  \Gscale@div\@tempa{\textheight}{\dimexpr\ht\pandoc@box+\dp\pandoc@box\relax}%
  \Gscale@div\@tempb{\linewidth}{\wd\pandoc@box}%
  \ifdim\@tempb\p@<\@tempa\p@\let\@tempa\@tempb\fi
  \ifdim\@tempa\p@<\p@\scalebox{\@tempa}{\usebox\pandoc@box}%
  \else\usebox{\pandoc@box}%
  \fi%
}
\def\fps@figure{htbp}
\makeatother
\ifLuaTeX
\usepackage[bidi=basic,shorthands=off]{babel}
\else
\usepackage[bidi=default,shorthands=off]{babel}
\fi
\ifLuaTeX
  \usepackage{selnolig} 
\fi
\providecommand{\tightlist}{%
  \setlength{\itemsep}{0pt}\setlength{\parskip}{0pt}}
\usepackage{bookmark}
\IfFileExists{xurl.sty}{\usepackage{xurl}}{} 
\hypersetup{
  pdftitle={Weird Machine Compositors: Exploiting AI Orchestration at the Expression Layer},
  pdfauthor={Eilon Cohen, Pillar Security; Ariel Fogel, Pillar Security},
  pdflang={en-US},
  colorlinks=true,
  linkcolor={Maroon},
  filecolor={Maroon},
  citecolor={Blue},
  urlcolor={Blue},
  pdfcreator={LaTeX via pandoc}}

\title{Weird Machine Compositors: Exploiting AI Orchestration at the
Expression Layer}
\author{Eilon Cohen\\Pillar Security \and Ariel Fogel\\Pillar Security}
\date{September 2026}

\input{preamble.tex}

\usepackage[numbers,sort&compress]{natbib}
\begin{document}
\maketitle

\begin{abstract}

Orchestration platforms secure user-provided expressions through
enumerate and block sandboxing: AST rewriting, runtime property
blocklists, template sandbox environments. We demonstrate that these
sandboxes are weird machines \cite{ref1} whose instruction
set is the underlying language specification, and that the enumerate and
block approach is unfixable, following the same trajectory that led to
the deprecation of past sandboxing technologies such as Java's
SecurityManager and vm2.

We validate this claim through three rounds of escalating bypasses
against n8n's expression sandbox (three CVEs, two CVSS 9.4, one
unauthenticated), and frame these findings within a broader pattern of
sandbox failures across the orchestration products category. We identify
a trust laundering pattern where orchestration pipelines and
applications move attacker controlled input from untrusted to fully
credentialed through transformations that strip taint at each level.
AI-assisted enumeration accelerates the discovery of these coverage
gaps, compressing the timeline between a sandbox's deployment and its
compromise.

We provide an AST coverage analysis methodology, an accompanying
open-source tool, and a defensive playbook that includes policy
inversion (allowlist over blocklist) as a structural mitigation.

\end{abstract}

\section{Introduction}\label{introduction}

\subsection{The Attack Scenario}\label{the-attack-scenario}

A company has mature AI security guardrails and controls: input and
output monitoring and security gates, model access controls. All
reporting green. An attacker submits a payload to a public ``Contact
Us'' form built on the company's workflow automation platform. Within
minutes, every stored credential is exfiltrated, every AI API call is
silently intercepted, and AI agent workflows execute attacker
instructions. This kind of compromise is upstream of every defense.

The platform was n8n, but the vulnerability class applies to most
orchestration tools that evaluate user expressions as code, including
agentic AI applications which function as orchestration platforms.

\subsection{Thesis}\label{thesis}

Expression sandboxes in orchestration platforms are weird machines
\cite{ref1}. The attacker's instruction set is the
language specification; the defender's blocklist is an incomplete
enumeration of that specification. When these weird machines compose
with each other, the result is emergent computation the individual
components were never designed to produce. The orchestration platform
launders trust between these stages, moving attacker-controlled input
from ``untrusted'' to ``fully credentialed'' through transformations
that strip taint at each hop.

\subsection{The Weird Machines Framework and Its
Extensions}\label{the-weird-machines-framework-and-its-extensions}

Bratus, Shapiro, and colleagues formalized a concept security
researchers had circled for decades: a \textbf{weird machine} is the
computational model an attacker actually exploits
\cite{ref1,ref2,ref3}.
It is the gap between what the programmer thinks the system computes and
what it actually computes. The attacker programs the weird machine by
providing inputs that trigger unintended computation.

Buffer overflows, format strings, and ROP (Return-Oriented Programming)
chains are all instances of weird machines
\cite{ref1,ref2,ref3}.
These examples operate at levels ranging from CPU instructions and
memory layout to ELF metadata and library internals, where the weird
machine's instruction set changes infrequently relative to the language
specifications that define expression engine instruction sets. We apply
the framework to expression engines in orchestration platforms and, in
doing so, identify three properties that require extending the framework
beyond its original low-level domain.

\textbf{Dynamic instruction sets.} In classical weird machines, the
instruction set is fixed: the CPU architecture does not change between
the attacker's analysis and exploitation. Expression engine weird
machines have instruction sets that grow with the language
specification. Each TC39 proposal introduces new AST node types that the
sandbox has not accounted for. The defender's problem is not to
enumerate a fixed set but to track a moving target. This is why
enumerate-and-block fails structurally rather than just practically: the
coverage gap widens with every specification revision, even if the
defender patches every known bypass.

\textbf{Compositional weird machines.} The Bratus literature describes
individual weird machines: a single program with a gap between intended
and actual computation. Orchestration platforms produce weird machines
through composition: two individually correct evaluation stages that,
when chained, create a weird machine neither produces alone. The Form
Node double-evaluation (Section 4.1) demonstrates this: neither
expression evaluation pass is vulnerable in isolation; their composition
produces RCE. The compositional weird machine's instruction set is the
set of inputs that chain through both stages: inputs that survive the
first stage and are recognized as executable by the second. This
instruction set exists only in the composition; inspecting either stage
alone will not reveal it.

\textbf{Trust laundering as the compositional mechanism.} In classical
weird machines, the attacker programs a single computational substrate.
In orchestration weird machines, the attacker's input crosses processing
boundaries where taint is stripped. We call this trust laundering: each
stage assumes its input is as trustworthy as its own context, so
attacker-controlled data gains authority at each hop without any
explicit trust decision. Trust laundering is what makes composition
exploitable in practice, because it is the mechanism by which the output
of one weird machine becomes trusted input to the next.

\subsection{Scope and Prior Art}\label{scope-and-prior-art}

This paper covers expression injection as a vulnerability class, the
weird machines framework applied to orchestration, and the emergent
attack surfaces created when expression engines compose with each other.
Viewing sandboxes as weird machines shifts the defensive paradigm from
``patching bugs'' to ``eliminating unintended instruction sets,'' which
is why the framework is necessary to understand why these sandboxes fail
structurally rather than incidentally.

\textbf{Prior art we build on:} - Bratus, Shapiro et al.~formalized the
weird machines framework
\cite{ref1,ref2,ref3}
- Poutievski and Ilgayev identified \texttt{\$\{\{\ \}\}} injection in
GitHub Actions \cite{ref10,ref11} -
vm2 was deprecated in 2023 after repeated sandbox escapes; resurrected
and broken again in 2026 (CVE-2026-22709) \cite{ref9} -
Java's SecurityManager was deprecated in JEP 411 after years of
escape-patch-escape cycles \cite{ref8} - Multiple
independent researchers discovered n8n sandbox escapes during the same
period: CVE-2025-68613 \cite{ref7} (discovered by
@fatihhcelik and @yuvalo1212), JFrog's CVE-2026-1470 and CVE-2026-0863
\cite{ref18}, and additional bypasses acknowledged in
n8n's v2.4.0 release notes. Each used different weird machine
instructions; all exploited the same structural gap - Sandbox failures
across Jinja2, Retool, and expr-eval confirm the vulnerability class
extends beyond n8n (Section 2.3)

\textbf{Our contributions:} 1. Proof that AST-based sandboxing cannot
serve as a security boundary, demonstrated through three rounds of
escalating bypasses against n8n and corroborated by independent findings
across platforms 2. Three extensions to the weird machines framework for
orchestration-layer systems: - \emph{Dynamic instruction sets}: the
attacker's instruction set grows with the language specification, unlike
the fixed instruction sets of classical weird machines -
\emph{Compositional weird machines}: individually correct stages compose
into emergent weird machines that neither produces alone - \emph{Trust
laundering}: taint is stripped at each processing boundary, enabling the
composition by allowing the output of one weird machine to become
trusted input to the next 3. An end-to-end kill-chain from
unauthenticated form submission to persistent credential interception,
demonstrating trust laundering across a full orchestration pipeline 4.
Policy inversion as a structural mitigation: replacing
enumerate-and-block with minimal-allowlist sandboxing 5. An AST coverage
analysis methodology, an accompanying open-source tool (to be released
with this paper), and a defensive playbook

\section{Threat Model and Expression
Injection}\label{threat-model-and-expression-injection}

\subsection{Attacker Model}\label{attacker-model}

Orchestration platforms create two classes of expression injection
attacker:

\textbf{Authenticated attacker.} Any user with workflow edit access. No
admin privileges required. In a multi-tenant deployment, this is any
tenant. The attacker creates or modifies workflows containing
expressions, the intended use of the platform. The attacker's capability
is indistinguishable from normal workflow authoring. In n8n, this
profile maps to CVE-2026-25049 and CVE-2026-27577.

\textbf{Unauthenticated attacker.} Any network-reachable actor. No
account, no authentication. The attacker's only requirement is a URL to
a public-facing endpoint that evaluates expressions with user-controlled
input. This profile exists on any platform that exposes
expression-evaluated content through public endpoints (forms, webhooks,
API responses). In n8n, this maps to CVE-2026-27493: Form endpoints
(\texttt{/\allowbreak{}form/\allowbreak{}*}) are unauthenticated by design. These endpoints are
discoverable via search engines and certificate transparency logs, and
the \texttt{/\allowbreak{}form/\allowbreak{}*} URL pattern is consistent across deployments.

\subsection{Deployment Model and Trust
Boundaries}\label{deployment-model-and-trust-boundaries}

Orchestration platforms share a common architectural pattern: the
expression engine, the credential vault, and the workflow execution
runtime coexist in a single process or trust domain. A sandbox escape in
the expression engine is a credential vault breach.

\input{trust-boundaries.tex}

\textbf{n8n instance.} n8n is used by over 230,000 organizations with
nearly 200 million Docker Hub pulls. It stores credentials for every
connected system (AWS keys, database passwords, OAuth tokens, AI
provider API keys) encrypted with a single environment variable. RCE on
the n8n process yields this key and, through it, every stored
credential. Retool disclosed the same architectural flaw in July 2025:
sandbox and encryption key sharing a process
\cite{ref21}.

When the pipeline includes multiple processing stages,
attacker-controlled input flows through multiple trust boundaries, each
transformation stripping taint, until it drives credentialed actions
(Section 4, trust laundering).

\subsection{Expression Injection as a Vulnerability
Class}\label{expression-injection-as-a-vulnerability-class}

Expression injection belongs to a class of vulnerabilities where
platforms evaluate user-supplied data as code. CI/CD pipelines had their
version with \texttt{\$\{\{\ \}\}} injection
\cite{ref10,ref11}. Orchestration
platforms have theirs with expression injection. The syntax varies; the
structural pattern is identical:

{\def\LTcaptype{none} 
\begin{longtable}[]{@{}
  >{\raggedright\arraybackslash}p{(\linewidth - 6\tabcolsep) * \real{0.1786}}
  >{\raggedright\arraybackslash}p{(\linewidth - 6\tabcolsep) * \real{0.1429}}
  >{\raggedright\arraybackslash}p{(\linewidth - 6\tabcolsep) * \real{0.3036}}
  >{\raggedright\arraybackslash}p{(\linewidth - 6\tabcolsep) * \real{0.3750}}@{}}
\toprule\noalign{}
\begin{minipage}[b]{\linewidth}\raggedright
Platform
\end{minipage} & \begin{minipage}[b]{\linewidth}\raggedright
Syntax
\end{minipage} & \begin{minipage}[b]{\linewidth}\raggedright
Sandbox Approach
\end{minipage} & \begin{minipage}[b]{\linewidth}\raggedright
Credential Proximity
\end{minipage} \\
\midrule\noalign{}
\endhead
\bottomrule\noalign{}
\endlastfoot
n8n & \texttt{=\{\{\ \}\}} & AST rewriting & Same process \\
Airflow & \texttt{\{\{\ \}\}} (Jinja) & Runtime attr blocklist & Same
process (Fernet key) \\
GitHub Actions & \texttt{\$\{\{\ \}\}} & None (shell expansion) &
Secrets in env vars \\
Retool & \texttt{\{\{\ \}\}} & In-process sandbox & Same process (enc
key) \\
Zapier & Code blocks & AWS Lambda isolation & Shared Lambda per
account \\
\end{longtable}
}

Each platform takes user-supplied data and evaluates it in a context
where it becomes code. Sandbox failures across the category confirm the
pattern:

\begin{itemize}
\tightlist
\item
  \textbf{Jinja2} (used by Apache Airflow and other orchestration
  tools): \texttt{Sandboxed\allowbreak{}Environment} underwent its own
  escape-patch-escape cycle. CVE-2024-56326 (indirect
  \texttt{str.\allowbreak{}format} access) was patched, then CVE-2025-27516 (bypass
  via \texttt{\textbar{}attr} filter) broke it again
  \cite{ref19,ref20}. Runtime
  attribute blocklist, same failure mode.
\item
  \textbf{Retool:} Disclosed an in-process sandbox escape (July 2025)
  with the same architectural flaw: sandbox and credential encryption
  key sharing a process \cite{ref21}.
\item
  \textbf{expr-eval:} A library purpose-built as a ``safe'' expression
  evaluator was broken (CVE-2025-12735), propagating through
  LangChain.js to thousands of AI applications
  \cite{ref22}.
\end{itemize}

CI/CD \texttt{\$\{\{\ \}\}} injection is now a well-understood
vulnerability class. Orchestration expression injection is the same
class in a different infrastructure layer. The syntax changes; the
pattern does not: user-supplied data evaluated as code in a context with
access to stored secrets.

\section{Empirical Proof: Three Rounds of Escalating
Bypasses}\label{empirical-proof-three-rounds-of-escalating-bypasses}

We validate the unfixability of enumerate-and-block sandboxing through
three rounds of escalating bypasses against n8n's expression engine. n8n
serves as the deep case study because it is representative of the
category (a custom AST-based sandbox protecting a credential vault) and
because the n8n security team's fast, competent responses produced the
strongest possible version of the defense for each round.

\subsection{Architecture of the n8n Expression
Sandbox}\label{architecture-of-the-n8n-expression-sandbox}

To demonstrate the gap between enumeration and specification, we analyze
n8n's \texttt{@n8n/\allowbreak{}tournament} \cite{ref13} as a
representative instance of AST-based sandboxing. The architecture is
typical: parse expressions into an AST, rewrite dangerous identifiers,
compile to executable code, wrap with runtime sanitizers. The
compilation pipeline (Figure 2):

\begin{enumerate}
\def\labelenumi{\arabic{enumi}.}
\tightlist
\item
  The expression string (e.g., \texttt{=\{\{\ \$json.name\ \}\}}) is
  parsed into an AST
\item
  A \texttt{Variable\allowbreak{}Polyfill} transformer walks the AST and rewrites
  bare identifiers (like \texttt{process}, \texttt{global}) into safe
  sandbox lookups (like \texttt{\_\_\_n8n\_data{[}"process"{]}})
\item
  The rewritten AST is compiled into a dynamically constructed
  JavaScript function
\item
  Runtime sanitizers (\texttt{Prototype\allowbreak{}Sanitizer},
  \texttt{Function\allowbreak{}This\allowbreak{}Sanitizer}, blocklists) wrap the execution
\end{enumerate}

\input{compilation-pipeline.tex}

The \texttt{Variable\allowbreak{}Polyfill}'s \texttt{visit\allowbreak{}Identifier} method is the
security-critical component. It checks each identifier's parent AST node
type against a switch statement to determine how to rewrite it:

\textbf{Listing 1:} Simplified \texttt{Variable\allowbreak{}Polyfill} switch
statement from \texttt{@n8n/\allowbreak{}tournament}.

\begin{Shaded}
\begin{Highlighting}[]
\ControlFlowTok{switch}\NormalTok{ (parent}\OperatorTok{.}\AttributeTok{type}\NormalTok{) \{}
    \ControlFlowTok{case} \StringTok{\textquotesingle{}MemberExpression\textquotesingle{}}\OperatorTok{:} \CommentTok{/* rewrite */} \ControlFlowTok{break}\OperatorTok{;}
    \ControlFlowTok{case} \StringTok{\textquotesingle{}CallExpression\textquotesingle{}}\OperatorTok{:}   \CommentTok{/* rewrite */} \ControlFlowTok{break}\OperatorTok{;}
    \ControlFlowTok{case} \StringTok{\textquotesingle{}BinaryExpression\textquotesingle{}}\OperatorTok{:} \CommentTok{/* rewrite */} \ControlFlowTok{break}\OperatorTok{;}
    \CommentTok{// ... 22 more cases ...}
    \ControlFlowTok{default}\OperatorTok{:}
        \ControlFlowTok{return} \FunctionTok{assertNever}\NormalTok{(parent)}\OperatorTok{;} \CommentTok{// no{-}op: returns true}
\NormalTok{\}}
\end{Highlighting}
\end{Shaded}

The full set of 25 handled parent node types:

\begin{Shaded}
\begin{Highlighting}[]
\NormalTok{MemberExpression    CallExpression       BinaryExpression}
\NormalTok{LogicalExpression   ConditionalExpression UnaryExpression}
\NormalTok{VariableDeclarator  AssignmentExpression Property}
\NormalTok{ArrayExpression     SequenceExpression   TemplateLiteral}
\NormalTok{TaggedTemplateExpression ReturnStatement IfStatement}
\NormalTok{SwitchCase          ThrowStatement       ForStatement}
\NormalTok{ForInStatement      ForOfStatement       WhileStatement}
\NormalTok{DoWhileStatement    ExpressionStatement  NewExpression}
\NormalTok{UpdateExpression}
\end{Highlighting}
\end{Shaded}

The ESTree specification defines 72 distinct AST node types
\cite{ref12,ref16}. The gap (47
unhandled types) is the weird machine's instruction set.

\subsection{Round 1: Template Literals and prepareStackTrace
(December 21,
2025)}\label{round-1-template-literals-and-preparestacktrace-december-21-2025}

\textbf{CVE-2026-25049 / GHSA-6cqr-8cfr-67f8}

The initial discovery chained three weird machine instructions:
\texttt{Tagged\allowbreak{}Template\allowbreak{}Expression} as a property accessor,
\texttt{Arrow\allowbreak{}Function\allowbreak{}Expression} as a \texttt{this} carrier, and
\texttt{Error.\allowbreak{}prepare\allowbreak{}Stack\allowbreak{}Trace} for code execution.

\textbf{Listing 2:} AST walkthrough of the Round 1 bypass.

\begin{Shaded}
\begin{Highlighting}[]

\NormalTok{  TaggedTemplateExpression}
\NormalTok{  +{-}{-} Identifier: target       tag position: treated as function call}
\NormalTok{      +{-}{-} TemplateLiteral      template: passes through as argument}
\NormalTok{          +{-}{-} expressions:}
\NormalTok{              +{-}{-} ArrowFunctionExpression   \textless{}{-}{-} parent.type = \textquotesingle{}ArrowFunction...\textquotesingle{}}
\NormalTok{                  +{-}{-} body:                     carries \textquotesingle{}this\textquotesingle{} reference}
\NormalTok{                      +{-}{-} CallExpression}
\NormalTok{                          +{-}{-} MemberExpression}
\NormalTok{                              +{-}{-} Error.prepareStackTrace}
\NormalTok{                                  (overrides stack trace formatting,}
\NormalTok{                                   executes attacker callback with}
\NormalTok{                                   access to real CallSite objects)}

\NormalTok{  The ArrowFunctionExpression carries a \textquotesingle{}this\textquotesingle{} binding the rewriter}
\NormalTok{  does not intercept. Error.prepareStackTrace provides a V8{-}specific}
\NormalTok{  code execution path outside normal function invocation.}
\end{Highlighting}
\end{Shaded}

Each node type evaded the rewriter because the specific combination
exploited gaps in how the switch statement handled parent-child
relationships.

n8n patched within 48 hours over the December holidays (PR \#23560). The
fix was correct for this specific vector.

Four days later, on December 24, a second bypass used
\texttt{Object.\allowbreak{}define\allowbreak{}Property} via \texttt{Call\allowbreak{}Expression} to manipulate
the prototype chain. Different weird machine instruction, same weird
machine. Two bypasses in four days against a fast, competent security
team. Both findings are covered by CVE-2026-25049
\cite{ref4}.

\subsection{Round 2: The Compilation-Stage Escape (February
2026)}\label{round-2-the-compilation-stage-escape-february-2026}

\textbf{CVE-2026-27577 / GHSA-vpcf-gvg4-6qwr}

In January, n8n shipped v2.4.0 with nine targeted security fixes:
\texttt{Prototype\allowbreak{}Sanitizer}, \texttt{Function\allowbreak{}This\allowbreak{}Sanitizer}, expanded
blocklists, additional AST node coverage. A thorough hardening effort
across multiple code paths.

In February, one expression bypassed all nine. Every fix was a runtime
check. The bypass operated at the compilation stage. The nine fixes
never executed. The payload uses \texttt{\{...process\}}, placing the
\texttt{process} identifier inside a \texttt{Spread\allowbreak{}Element} AST node.
The \texttt{Variable\allowbreak{}Polyfill}'s switch statement has no case for
\texttt{Spread\allowbreak{}Element}. It falls through to \texttt{default}.
\texttt{assert\allowbreak{}Never()} returns \texttt{true}, and no rewriting occurs.
The bare \texttt{process} identifier resolves to Node.js's real
\texttt{process} global. From there, \texttt{get\allowbreak{}Builtin\allowbreak{}Module} returns
the \texttt{child\_process} module, and arbitrary command execution
follows.

\textbf{Listing 3:} AST walkthrough of the SpreadElement bypass.

\begin{Shaded}
\begin{Highlighting}[]

\NormalTok{  ObjectExpression}
\NormalTok{  +{-}{-} SpreadElement          \textless{}{-}{-} parent.type = \textquotesingle{}SpreadElement\textquotesingle{}}
\NormalTok{      +{-}{-} Identifier: process   NOT in switch {-}{-}\textgreater{} default {-}{-}\textgreater{} no rewrite}
\NormalTok{                                 Bare \textquotesingle{}process\textquotesingle{} resolves to Node.js global}

\NormalTok{  Compare with handled case:}
\NormalTok{  MemberExpression}
\NormalTok{  +{-}{-} Identifier: process   IN switch {-}{-}\textgreater{} rewritten to \_\_\_n8n\_data["process"]}
\NormalTok{                             Resolves to undefined in sandbox}
\end{Highlighting}
\end{Shaded}

\textbf{The critical insight:} This bypass operates at the compilation
stage, before any runtime sanitizer runs. All nine v2.4.0 fixes are
runtime checks: \texttt{Prototype\allowbreak{}Sanitizer},
\texttt{Function\allowbreak{}This\allowbreak{}Sanitizer}, expanded blocklists, immutable
\texttt{\_\_sanitize}, destructuring protection. They never execute
because the identifier escapes during AST rewriting, before the compiled
function is even constructed.

{\def\LTcaptype{none} 
\begin{longtable}[]{@{}
  >{\raggedright\arraybackslash}p{(\linewidth - 2\tabcolsep) * \real{0.3000}}
  >{\raggedright\arraybackslash}p{(\linewidth - 2\tabcolsep) * \real{0.7000}}@{}}
\toprule\noalign{}
\begin{minipage}[b]{\linewidth}\raggedright
Defense Layer
\end{minipage} & \begin{minipage}[b]{\linewidth}\raggedright
Why It Fails Against SpreadElement
\end{minipage} \\
\midrule\noalign{}
\endhead
\bottomrule\noalign{}
\endlastfoot
\texttt{Variable\allowbreak{}Polyfill} & \texttt{Spread\allowbreak{}Element} not in switch; falls
to no-op default \\
\texttt{This\allowbreak{}Sanitizer} & Blocks \texttt{global\allowbreak{}This}, not
\texttt{process} \\
\texttt{Prototype\allowbreak{}Sanitizer} & \texttt{get\allowbreak{}Builtin\allowbreak{}Module} not in blocked
properties \\
\texttt{Function\allowbreak{}This\allowbreak{}Sanitizer} & No \texttt{this} access used \\
\texttt{\_\_sanitize} (runtime) & Only wraps computed
\texttt{obj{[}x{]}} access; all accesses here are dot notation \\
All 9 v2.4.0 fixes & Runtime-layer fixes; bypass is at compilation
stage \\
\end{longtable}
}

\subsection{The Fix-Bypass Cycle as Structural
Proof}\label{the-fix-bypass-cycle-as-structural-proof}

The three rounds constitute empirical evidence that AST enumeration
cannot work:

{\def\LTcaptype{none} 
\begin{longtable}[]{@{}
  >{\raggedright\arraybackslash}p{(\linewidth - 6\tabcolsep) * \real{0.1045}}
  >{\raggedright\arraybackslash}p{(\linewidth - 6\tabcolsep) * \real{0.0896}}
  >{\raggedright\arraybackslash}p{(\linewidth - 6\tabcolsep) * \real{0.3881}}
  >{\raggedright\arraybackslash}p{(\linewidth - 6\tabcolsep) * \real{0.4179}}@{}}
\toprule\noalign{}
\begin{minipage}[b]{\linewidth}\raggedright
Round
\end{minipage} & \begin{minipage}[b]{\linewidth}\raggedright
Date
\end{minipage} & \begin{minipage}[b]{\linewidth}\raggedright
Weird Machine Instruction
\end{minipage} & \begin{minipage}[b]{\linewidth}\raggedright
What the Enumeration Missed
\end{minipage} \\
\midrule\noalign{}
\endhead
\bottomrule\noalign{}
\endlastfoot
1 & Dec 21 & \texttt{Tagged\allowbreak{}Template\allowbreak{}Expression} +
\texttt{Arrow\allowbreak{}Function\allowbreak{}Expression} (CVE-2026-25049) & Template literals as
function calls \\
1.5 & Dec 24 & \texttt{Object.\allowbreak{}define\allowbreak{}Property} via
\texttt{Call\allowbreak{}Expression} (covered by CVE-2026-25049) & Property
descriptor manipulation \\
2 & Feb 7 & \texttt{Spread\allowbreak{}Element} at compilation stage (CVE-2026-27577)
& Spread syntax bypassing identifier rewriting entirely \\
\end{longtable}
}

Each fix was correct. Each was rendered irrelevant by the next bypass,
as shown in Figure 3.

\input{fix-bypass-cycle.tex}

{\def\LTcaptype{none} 
\begin{longtable}[]{@{}
  >{\raggedright\arraybackslash}p{(\linewidth - 4\tabcolsep) * \real{0.3000}}
  >{\raggedright\arraybackslash}p{(\linewidth - 4\tabcolsep) * \real{0.3500}}
  >{\raggedright\arraybackslash}p{(\linewidth - 4\tabcolsep) * \real{0.3500}}@{}}
\toprule\noalign{}
\begin{minipage}[b]{\linewidth}\raggedright
Date
\end{minipage} & \begin{minipage}[b]{\linewidth}\raggedright
Actor
\end{minipage} & \begin{minipage}[b]{\linewidth}\raggedright
Event
\end{minipage} \\
\midrule\noalign{}
\endhead
\bottomrule\noalign{}
\endlastfoot
Dec 21 & Pillar & Round 1 discovered and reported (template literal
chain) \\
Dec 23 & n8n & Patch PR \#23560 shipped (48-hour holiday response) \\
Dec 24 & Pillar & Round 1.5 discovered and reported (ODP bypass) \\
Jan & n8n & v2.4.0 released with 9 targeted security fixes \\
Jan & n8n & v2.4.0 acknowledges additional bypasses from other
reporters \\
Jan & JFrog & CVE-2026-1470 (\texttt{with} statement) and CVE-2026-0863
(Python introspection) published \\
Feb 3 & Joint & Public disclosure of Round 1 findings
\cite{ref15} \\
Feb 7 & Pillar & Round 2 discovered and reported (SpreadElement);
bypasses all 9 v2.4.0 fixes \\
Feb 25 & n8n & Patch PR \#26214 shipped \\
\end{longtable}
}

Multiple independent bypass discoveries across several research teams in
three months.

\textbf{Independent corroboration.} n8n's expression sandbox attracted
multiple independent researchers during the same period. CVE-2025-68613
(CVSS 9.9) was disclosed separately \cite{ref7}. JFrog
published two additional escapes: CVE-2026-1470 used the \texttt{with}
statement to resolve \texttt{constructor} to the \texttt{Function}
constructor, and CVE-2026-0863 exploited Python's
\texttt{Attribute\allowbreak{}Error} introspection to bypass the Python Code node's
AST sandbox \cite{ref18}. n8n's v2.4.0 release notes
acknowledged additional bypasses from other reporters. Each researcher
used different weird machine instructions; all exploited the same
structural gap. The volume of independent discoveries across multiple
teams, using unrelated vectors, is itself evidence that the
vulnerability class is inherent to the approach.

The trajectory matches two known precedents:

\textbf{Java SecurityManager} \cite{ref8}: Sun/Oracle
constrained Java through permission checks over a rich instruction set.
Years of escape-patch-escape. Oracle deprecated SecurityManager in Java
17 because the enumeration approach could not be made sound.

\textbf{vm2} \cite{ref9}: The most popular Node.js
sandbox. Deprecated in 2023 after repeated escapes. Resurrected in 2026.
Broken again (CVE-2026-22709). Same arc.

n8n's \texttt{@n8n/\allowbreak{}tournament} sandbox is on the same trajectory. The
defender maintains a list of dangerous patterns. The ECMAScript
specification grows with every TC39 proposal. The defender's enumeration
falls further behind with every spec revision.

\section{Composition and Trust
Laundering}\label{composition-and-trust-laundering}

\subsection{Composition Creates New Attack
Surfaces}\label{composition-creates-new-attack-surfaces}

When two expression evaluation stages feed into each other, they compose
into a weird machine that neither stage's designer anticipated. Each
stage is individually correct; the composition produces unintended
computation.

\textbf{CVE-2026-27493 / GHSA-75g8-rv7v-32f7: Form Node
Double-Evaluation}

We demonstrate this with n8n's Form nodes, which create public-facing,
unauthenticated endpoints. A business user builds a multi-step form
where step 2 displays ``Thank you, \{\{Name\}\}!''. No code required, no
security concern apparent. The vulnerability: two expression evaluation
passes with user input flowing between them unsanitized.

\textbf{Pass 1} (\texttt{get\allowbreak{}Node\allowbreak{}Parameter}, Form.node.ts:374): Resolves
configured expressions. The HTML field
\texttt{=\{\{\ \$input.first().json{[}"Name"{]}\ \}\}} evaluates to
whatever the user submitted. If the user typed expression syntax, it
passes through as a literal string.

\textbf{Pass 2} (\texttt{prepare\allowbreak{}Form\allowbreak{}Fields}, utils.ts:119-139): Scans
the already-resolved HTML for \texttt{\{\{\ \}\}} patterns using
\texttt{get\allowbreak{}Resolvables()} and evaluates each one via
\texttt{evaluate\allowbreak{}Expression()}. It finds the attacker's payload. It
evaluates it.

{\def\LTcaptype{none} 
\begin{longtable}[]{@{}
  >{\raggedright\arraybackslash}p{(\linewidth - 6\tabcolsep) * \real{0.1795}}
  >{\raggedright\arraybackslash}p{(\linewidth - 6\tabcolsep) * \real{0.2564}}
  >{\raggedright\arraybackslash}p{(\linewidth - 6\tabcolsep) * \real{0.3590}}
  >{\raggedright\arraybackslash}p{(\linewidth - 6\tabcolsep) * \real{0.2051}}@{}}
\toprule\noalign{}
\begin{minipage}[b]{\linewidth}\raggedright
Stage
\end{minipage} & \begin{minipage}[b]{\linewidth}\raggedright
Function
\end{minipage} & \begin{minipage}[b]{\linewidth}\raggedright
What Happens
\end{minipage} & \begin{minipage}[b]{\linewidth}\raggedright
Result
\end{minipage} \\
\midrule\noalign{}
\endhead
\bottomrule\noalign{}
\endlastfoot
Pass 1 & getNodeParameter & Resolves configured expressions; attacker
input passes through as literal string & Attacker payload embedded in
resolved HTML \\
Pass 2 & prepareFormFields & Scans resolved HTML for expression
patterns; evaluates attacker payload as code & RCE or data leak \\
\end{longtable}
}

Neither pass is vulnerable alone. The composition produces RCE.

This parallels Return-Oriented Programming: individual gadgets are
benign, but their composition achieves arbitrary computation. The Form
Node is an evaluation chain, two ROP gadgets in sequence. The
difference: no authentication is required. Form endpoints are public by
design.

\textbf{The compositional instruction set.} The composition produces a
weird machine whose instruction set exists only in the interaction
between stages. Pass 1's instruction set is the set of expressions the
platform's configured fields can resolve; it treats user-submitted form
data as literal strings, not as expressions. Pass 2's instruction set is
the set of \texttt{\{\{\ \}\}} patterns that \texttt{get\allowbreak{}Resolvables()}
recognizes in already-resolved HTML. Neither instruction set alone
produces attacker-controlled code execution. Pass 1 cannot evaluate
attacker input as code because it treats form data as literals. Pass 2
cannot evaluate attacker input as code because it only scans HTML that
has already been resolved by Pass 1, and under normal operation that
HTML contains no attacker-controlled expression patterns. The
compositional instruction set is the set of form inputs that Pass 1
embeds as literal strings into HTML \emph{and} that Pass 2 recognizes as
expression patterns in the resolved output. This set is empty when
either stage is analyzed alone; it exists only when the stages are
chained.

Auditing Pass 1 reveals no vulnerability. Auditing Pass 2 reveals no
vulnerability. Auditing the composition reveals RCE.

\textbf{The double-evaluation is independently dangerous.} Even with a
perfect sandbox, evaluating attacker-controlled expressions leaks
workflow data, environment variables, and execution context. The
expression injection is the vulnerability; the sandbox escape is one
payload. Patching the sandbox does not fix the injection.

\subsection{Trust Laundering}\label{trust-laundering}

In most orchestration platforms, pipelines implicitly escalate trust at
each processing step. No step makes an explicit trust decision; each
step assumes its input is as trustworthy as its own context. We call
this \textbf{trust laundering}. Attacker-controlled input is transformed
at each hop, with taint stripped at each boundary, until it drives fully
credentialed or unintended actions.

To illustrate the full pipeline, consider a typical workflow: a
public-facing form feeds a code node that enriches the request, which
feeds a data transformation node that formats a response, which feeds an
external communication node that sends the reply using stored
credentials. The pipeline spans multiple nodes and multiple trust
levels, but at no boundary does any stage make an explicit trust
decision:

{\def\LTcaptype{none} 
\begin{longtable}[]{@{}
  >{\raggedright\arraybackslash}p{(\linewidth - 4\tabcolsep) * \real{0.1765}}
  >{\raggedright\arraybackslash}p{(\linewidth - 4\tabcolsep) * \real{0.1765}}
  >{\raggedright\arraybackslash}p{(\linewidth - 4\tabcolsep) * \real{0.6471}}@{}}
\toprule\noalign{}
\begin{minipage}[b]{\linewidth}\raggedright
Node
\end{minipage} & \begin{minipage}[b]{\linewidth}\raggedright
Role
\end{minipage} & \begin{minipage}[b]{\linewidth}\raggedright
Trust Level After Node
\end{minipage} \\
\midrule\noalign{}
\endhead
\bottomrule\noalign{}
\endlastfoot
Form Trigger (\texttt{/\allowbreak{}form/\allowbreak{}support}) & Receives name, email, query &
\textbf{Untrusted} user input \\
Code Node & Evaluates expressions to enrich request & ``Workflow data''
(taint stripped) \\
Set Node & Formats response from enriched context & ``Processed data''
(taint stripped) \\
External Communication Node & Sends reply using stored credentials &
\textbf{Fully trusted} API call \\
\end{longtable}
}

\input{trust-laundering-pipeline.tex}

An attacker who achieves RCE through expression injection (Section 3)
can modify this pipeline at any stage: alter the Code Node's logic, swap
a credential's base URL to route traffic through an attacker proxy, or
modify the Gmail Node's recipient to BCC all responses to an external
address. Each node behaves as designed, but the composition is where the
attacker operates, and each transformation strips taint from the
attacker's input.

\subsection{The Kill-Chain: Expression Injection to Persistent
Credential
Interception}\label{the-kill-chain-expression-injection-to-persistent-credential-interception}

Section 4.1 demonstrated two-stage deterministic composition: two
individually correct evaluation passes that produce RCE when chained.
The following kill-chain demonstrates trust laundering across a full
pipeline: expression injection escalates through credential extraction
and base URL tampering to persistent, silent interception of an
organization's credentialed API calls. We validate this with an
end-to-end attack against n8n \cite{ref14}, building on
the credential extraction and base URL swap methodology from
CVE-2025-68613 \cite{ref7}. The six steps:

\begin{enumerate}
\def\labelenumi{\arabic{enumi}.}
\item
  Expression injection achieves RCE (CVE-2026-27577 or CVE-2026-27493)
\item
  RCE reads the encryption key from the process environment
\item
  The encryption key decrypts every stored credential (OpenAI,
  Anthropic, AWS, etc.)
\item
  The attacker modifies an AI provider's base URL in n8n's credential
  store to point to an attacker-controlled proxy

  n8n stores credentials in a \texttt{credentials\_entity} database
  table as base64-encoded blobs encrypted with AES-256-CBC in
  OpenSSL-compatible format: a \texttt{Salted\_\_} prefix, 8-byte salt,
  and key derived via \texttt{EVP\_Bytes\allowbreak{}To\allowbreak{}Key} with MD5 (a deprecated
  KDF that provides no stretching or brute-force resistance beyond the
  entropy of the input key). With the encryption key from step 2, the
  attacker's RCE payload queries the database, decodes and decrypts each
  credential blob, parses the resulting JSON, modifies the
  \texttt{base\allowbreak{}Url} field, re-encrypts, and writes the row back
  \cite{ref14}
\item
  All subsequent API calls to that provider route through the proxy
  transparently. The proxy is a simple HTTPS reverse proxy that
  terminates TLS with its own certificate; n8n does not validate the
  base URL's TLS certificate chain against any pinned or expected CA.
  The proxy forwards requests to the real provider, captures API keys
  from \texttt{Authorization} headers and full request/response
  payloads, and returns unmodified responses. Latency added by the proxy
  hop is indistinguishable from normal API variance. Implementation
  details and the proxy code are in the n8n-poc repository
  \cite{ref14}
\item
  The n8n workflow completes normally. All monitoring reports green. The
  attack persists across n8n restarts because the modified credential is
  stored in the database
\end{enumerate}

Each individual stage behaves as designed. The composition achieves
persistent, silent interception of an organization's credentialed
infrastructure. The database modification in step 4 is itself trust
laundering: the attacker uses the expression engine RCE to permanently
poison the credential store, ensuring all future API calls operate on
behalf of the attacker. Downstream monitoring cannot detect it because
the compromise is upstream of every check. Figure 5 illustrates the full
chain.

\input{kill-chain.tex}

\section{AST Coverage Methodology and
Tool}\label{ast-coverage-methodology-and-tool}

\subsection{Quantifying the Weird
Machine}\label{quantifying-the-weird-machine}

The weird machines framework gives us a way to quantify expression
engine attack surfaces. If the weird machine's instruction set is the
set of AST node types the sandbox does not handle, we can measure it.

We developed a five-step methodology (detailed in Appendix C):

\begin{enumerate}
\def\labelenumi{\arabic{enumi}.}
\tightlist
\item
  \textbf{Extract} the expression engine's AST rewriting logic
\item
  \textbf{Enumerate} the node types it handles
\item
  \textbf{Compare} against the full language specification's node types
  \cite{ref12,ref16}
\item
  \textbf{Classify} unhandled types by exploitation potential:

  \begin{itemize}
  \tightlist
  \item
    \emph{Critical}: can carry a bare global reference that escapes
    rewriting (direct RCE path)
  \item
    \emph{High}: can alter control flow or scope to bypass runtime
    sanitizers
  \item
    \emph{Medium}: can access properties or invoke behavior not covered
    by blocklists
  \item
    \emph{Low}: syntactically valid but unlikely to carry exploitable
    semantics
  \end{itemize}
\item
  \textbf{Score} the coverage gap
\end{enumerate}

\subsection{Results for n8n's
tournament}\label{results-for-n8ns-tournament}

Applied to \texttt{@n8n/\allowbreak{}tournament} \cite{ref13}:

{\def\LTcaptype{none} 
\begin{longtable}[]{@{}ll@{}}
\toprule\noalign{}
Metric & Value \\
\midrule\noalign{}
\endhead
\bottomrule\noalign{}
\endlastfoot
Handled node types & 25 \\
ESTree spec node types & 72 \\
Unhandled (weird machine instruction set) & 47 \\
\textbf{Coverage ratio} & \textbf{35\%} \\
\end{longtable}
}

\input{ast-coverage-gap.tex}

\textbf{Confirmed exploitable (now patched):}

\begin{itemize}
\tightlist
\item
  \texttt{Spread\allowbreak{}Element}: CVE-2026-27577 (CVSS 9.4). Reported, patched
  in PR \#26214.
\end{itemize}

\textbf{Unhandled types requiring further analysis.} The remaining 46
unhandled node types have not been individually tested for
exploitability. We classify them by whether the node type can
syntactically carry a bare global reference or alter scope:

\begin{itemize}
\tightlist
\item
  \emph{Structurally similar to SpreadElement} (can carry identifiers in
  parent positions the switch does not handle): \texttt{Rest\allowbreak{}Element}
\item
  \emph{Can introduce scope or control flow changes}:
  \texttt{Yield\allowbreak{}Expression}, \texttt{Await\allowbreak{}Expression},
  \texttt{Class\allowbreak{}Expression}, \texttt{Import\allowbreak{}Expression},
  \texttt{Meta\allowbreak{}Property}
\end{itemize}

These classifications reflect static AST analysis, not confirmed
exploitation; responsible assessment requires coordinated testing for
each type.

\subsection{The Tool}\label{the-tool}

We are building an open-source AST coverage analyzer that automates this
methodology. The reference implementation targets
\texttt{@n8n/\allowbreak{}tournament} and will be released as open source. The
design:

\begin{itemize}
\tightlist
\item
  Parses an expression engine's AST rewriting logic and extracts handled
  node types
\item
  Maps handled types against the full ECMAScript AST specification
  (ESTree \cite{ref16} / Babel AST)
\item
  Reports unhandled node types with severity ratings based on
  exploitation potential
\item
  Generates a coverage report with the ratio of defended vs.~undefended
  instruction set
\item
  Supports pluggable parsers for different expression engine
  architectures
\end{itemize}

The manual analysis in Section 5.2 validates the methodology; the tool
encodes it as a repeatable, automated workflow. The pluggable parser
interface will enable analysis of Jinja2-based sandboxes and custom
expression engines beyond the \texttt{@n8n/\allowbreak{}tournament} reference
implementation.

\subsection{AI-Assisted Enumeration of Coverage
Gaps}\label{ai-assisted-enumeration-of-coverage-gaps}

The coverage gap identified in Section 5.2 (47 unhandled node types)
defines the weird machine's instruction set, but static classification
alone does not confirm exploitability. To test unhandled types
systematically, we used an LLM to generate candidate bypass expressions
for each unhandled node type in the ESTree specification, then tested
each against the sandbox. The SpreadElement bypass (CVE-2026-27577) was
identified through this process: systematic analysis of the gap between
the switch statement and the specification, with AI tooling generating
and refining test cases for each unhandled type.

Enumerate-and-block sandboxes have historically relied on the obscurity
of their coverage gaps as a practical mitigating factor: even if the
sandbox is theoretically incomplete, finding an exploitable gap requires
manual analysis of the specification and the sandbox implementation.
AI-assisted enumeration eliminates this reliance. A 35\% coverage ratio
means 47 potential bypass vectors. Generating and testing candidate
expressions for all 47 is a matter of hours, not months.

\section{Defensive Playbook}\label{defensive-playbook}

\subsection{What Does Not Work}\label{what-does-not-work}

These defensive failures are properties of the approach, not of any
specific platform.

\textbf{AST rewriting as a security boundary.} Enumerating dangerous AST
node types against a growing language specification cannot keep up. Each
TC39 proposal introduces new syntax and new node types the blocklist has
not accounted for. Our three rounds of n8n bypasses demonstrate this;
Java's SecurityManager and vm2 demonstrate it historically.

\textbf{Runtime sanitizers alone.} Compilation-stage escapes bypass all
runtime sanitizers. The SpreadElement bypass (CVE-2026-27577) escaped
during AST rewriting; nine runtime fixes never executed. Any sandbox
that rewrites at compile time and sanitizes at runtime has this gap.

\textbf{Multi-stage evaluation.} When the output of one expression
evaluation feeds into another, the composition creates injection vectors
even without a sandbox escape. Any platform where resolved output is
scanned for expression patterns and re-evaluated has this exposure.

\subsection{What Works}\label{what-works}

\textbf{Policy inversion: from blocklist to allowlist.} The fundamental
flaw of enumerate-and-block is that it defines the sandbox by what it
excludes. Defenders must enumerate every dangerous AST node type, a set
that grows with every specification revision. We believe that policy
inversion reverses this: define the sandbox by the minimal set of AST
node types required to support the intended workflow functionality, and
reject everything else by default.

This is the same shift that transformed other domains: default-deny
firewall rules replaced default-allow; Content Security Policy replaced
unrestricted inline script execution; seccomp-bpf in container runtimes
replaced unrestricted syscall access. In each case, the
enumerate-and-block approach failed because the attack surface grew
faster than the blocklist.

Policy inversion does not eliminate the need for runtime isolation (see
below). An attacker who finds an exploitable combination within the
allowed node types can still escape. But it reduces the weird machine's
instruction set to a small, auditable surface, and it ensures that
specification growth does not automatically expand the attack surface in
each new version of the specification language.

\textbf{Runtime isolation.} V8 isolates, Deno sandboxes, Cloudflare
Workers, or separate processes with IPC. The expression engine should
physically not be able to access the host process, regardless of syntax.
Process-level isolation succeeds where AST rewriting fails because it
shifts the security boundary from the language specification (infinite,
growing with every TC39 proposal) to the operating system kernel
(finite, stable, decades of hardening). AST rewriting, whether blocklist
or allowlist, is useful as defense-in-depth, but runtime isolation is
the minimum viable security boundary. As of March 2026, n8n still uses
the tournament sandbox with SpreadElement patched. The underlying
vulnerability class remains; only the known vectors are blocked. No
major orchestration platform has adopted process-level isolation for
expression evaluation. Zapier's Lambda-based isolation is the closest
existing model, though its initial sharing flaw (ZAPESCAPE
\cite{ref23}) shows that isolation boundaries require
careful design.

\textbf{Single-pass evaluation.} Review all multi-stage evaluation
patterns of user-controlled inputs, and eliminate what isn't strictly
necessary. If user data passes through one evaluation, it should not
pass through another, where this could be feasible without breaking
platform functionality. For instance, n8n's patch for CVE-2026-27493
removed the second evaluation pass from \texttt{prepare\allowbreak{}Form\allowbreak{}Fields}
entirely, preserving all Form Node features while eliminating the double
evaluation. The second pass was unnecessary as the first pass already
resolved all intended expressions.

\textbf{Process-level credential separation.} The expression engine
should not run in the same process as decrypted secrets. If the sandbox
fails, the secrets must still be unreachable to the attacker. This
requires architectural changes: separate processes for expression
evaluation and credential management, communicating through a narrow,
authenticated IPC interface.

\subsection{Immediate Actions for
Defenders}\label{immediate-actions-for-defenders}

\begin{enumerate}
\def\labelenumi{\arabic{enumi}.}
\item
  \textbf{Treat your orchestration platform as a credential vault.} Same
  audit cadence, same access controls, same monitoring as vaults. If
  your orchestration platform stores credentials, it is a credential
  vault.
\item
  \textbf{Eliminate multi-stage evaluation patterns.} Audit every node
  and component for double-evaluation: any path where resolved output is
  scanned for expression patterns and evaluated again.
\item
  \textbf{Audit expression sandbox coverage.} Apply the AST coverage
  methodology (Section 5) to determine what percentage of the language
  specification your sandbox handles. If the coverage ratio is below
  100\%, the sandbox is a weird machine.
\item
  \textbf{Segment credential access from expression evaluation.} At
  minimum, process-level isolation. Expression evaluation in an isolated
  process that cannot read the platform's encryption key or access the
  credential database.
\item
  \textbf{Rotate credentials assuming prior breach.} Any orchestration
  platform instance running a version with a known sandbox escape could
  have exposed its encryption key. Rotate all stored credentials.
\end{enumerate}

\section{Conclusion and Future
Work}\label{conclusion-and-future-work}

\subsection{Summary}\label{summary}

We have demonstrated that expression sandboxes in orchestration
platforms are weird machines whose instruction set is the language
specification. The three rounds of bypasses against n8n's progressively
hardened sandbox (three critical CVEs: CVSS 9.4, 9.4, 9.5; one requiring
no authentication) provide empirical proof that AST-based sandboxing
follows the same trajectory as other attempts to sandbox dynamic
languages.

Applying the weird machines framework to orchestration-layer systems
required three extensions. Expression engine weird machines have dynamic
instruction sets that grow with the language specification, not fixed as
the instruction sets of classical weird machines presented in the
original paper. The AST coverage analysis quantifies this gap, and the
gap widens with every specification revision. Beyond dynamic instruction
sets, orchestration platforms produce compositional weird machines:
individually correct processing stages that compose into emergent weird
machines neither produces alone, as demonstrated by the Form Node
double-evaluation. The mechanism that makes this composition exploitable
is trust laundering --- each stage strips taint from its input, so
attacker-controlled data gains authority at each hop without any
explicit trust decision. We demonstrated this through a full kill-chain
from unauthenticated form submission to persistent credential
interception.

We proposed policy inversion as a structural mitigation: replacing
enumerate-and-block with minimal-allowlist sandboxing, so that
specification growth does not automatically expand the attack surface.
Policy inversion complements runtime isolation; neither alone is
sufficient, but together they address the vulnerability class at both
the language layer and the process layer.

\subsection{Three Takeaways}\label{three-takeaways}

\begin{enumerate}
\def\labelenumi{\arabic{enumi}.}
\item
  \textbf{Expression injection is a vulnerability class.} Every platform
  that evaluates user-supplied expressions near stored credentials has
  the same exposure.
\item
  \textbf{AST enumeration cannot be a security boundary.} The fix-bypass
  cycle repeats because the defender's enumeration falls further behind
  the specification with every revision. Policy inversion (allowlist
  over blocklist) reduces the weird machine's instruction set; runtime
  isolation (V8 isolates, process-level sandboxing) is the minimum
  viable security boundary.
\item
  \textbf{The weird machines framework extends to orchestration.}
  Classical weird machines have fixed instruction sets and operate on a
  single computational substrate. Orchestration weird machines have
  dynamic instruction sets, compose across processing stages, and
  exploit trust laundering to move attacker-controlled input from
  untrusted to fully credentialed.
\end{enumerate}

\subsection{Future Work}\label{future-work}

\textbf{Cross-platform validation.} Our deep case study is n8n. But the
evidence that enumerate-and-block sandboxing cannot hold already extends
well beyond it:

{\def\LTcaptype{none} 
\begin{longtable}[]{@{}
  >{\raggedright\arraybackslash}p{(\linewidth - 4\tabcolsep) * \real{0.2703}}
  >{\raggedright\arraybackslash}p{(\linewidth - 4\tabcolsep) * \real{0.4595}}
  >{\raggedright\arraybackslash}p{(\linewidth - 4\tabcolsep) * \real{0.2703}}@{}}
\toprule\noalign{}
\begin{minipage}[b]{\linewidth}\raggedright
Platform
\end{minipage} & \begin{minipage}[b]{\linewidth}\raggedright
Sandbox Approach
\end{minipage} & \begin{minipage}[b]{\linewidth}\raggedright
Evidence
\end{minipage} \\
\midrule\noalign{}
\endhead
\bottomrule\noalign{}
\endlastfoot
n8n & AST rewriting & 6+ CVEs from 4+ independent teams in 3 months \\
Jinja2 (Airflow) & Runtime attr blocklist & Escape-patch-escape cycle \\
Retool & In-process sandbox & Sandbox escape, July 2025 \\
expr-eval (LangChain.js) & Allowlist evaluator & CVE-2025-12735,
thousands of apps \\
Zapier & AWS Lambda isolation & Isolation held; sharing model broke \\
\end{longtable}
}

\textbf{n8n:} 3 CVEs from Pillar
\cite{ref4,ref5,ref6},
2 from JFrog \cite{ref18}, CVE-2025-68613 from
independent researchers \cite{ref7}, plus additional
reports acknowledged in n8n's v2.4.0 release notes --- 6+ CVEs from 4+
independent teams in three months. \textbf{Jinja2:} CVE-2024-56326
patched, then CVE-2025-27516 bypassed it
\cite{ref19,ref20}. \textbf{Retool:}
Same architecture as n8n --- sandbox and encryption key sharing a
process \cite{ref21}. \textbf{expr-eval:} ``Safe''
evaluator broken, propagated through LangChain.js
\cite{ref22}. \textbf{Zapier:} ZAPESCAPE
\cite{ref23} exploited the sharing model (shared Lambda
per account), not the isolation mechanism.

Zapier is instructive as a partial counterexample: they chose process
isolation (Lambda) over language-level sandboxing. The vulnerability
(ZAPESCAPE) was in the \emph{sharing model} (shared Lambda per account),
not in the isolation mechanism itself. Once Zapier moved to per-user
Lambda instances, the issue was mitigated. This supports our defensive
recommendation: architectural isolation works; enumerate-and-block does
not.

The AST coverage analysis methodology generalizes to any platform with
an expression engine and a sandboxing layer. We provide the methodology
and the tooling. Each platform analyzed strengthens the evidence that
expression injection is a class. We invite the research community to
apply the analysis to other orchestration platforms and publish the
results.

\textbf{Compositional weird machines with stochastic components.} The
compositional extension developed in this paper covers deterministic
expression engines composing with each other. A natural next step is to
examine whether the extension holds when one component is stochastic,
specifically when an LLM processes natural language within the same
orchestration pipeline. Whether such composition produces emergent
attack surfaces beyond what either component creates alone is an open
question that the framework extensions established here provide a
foundation for investigating.

\subsection{Responsible Disclosure}\label{responsible-disclosure}

All vulnerabilities were reported to n8n through coordinated disclosure.
n8n responded within 24-48 hours to every report, including over the
December holidays. Patches are available for all findings. The full
disclosure timeline is in Appendix B.

n8n's security team was professional and responsive throughout. The
problems we describe are industry-wide. n8n's willingness to engage
seriously with each finding made this research possible.

\clearpage
\appendix
\section{CVE Details}\label{appendix-a-cve-details}

\subsection{Summary Table}\label{summary-table}

{\def\LTcaptype{none} 
\begin{longtable}[]{@{}lllll@{}}
\toprule\noalign{}
CVE & Round & Vector & CVSS & Auth \\
\midrule\noalign{}
\endhead
\bottomrule\noalign{}
\endlastfoot
CVE-2026-25049 & 1 & Template literals + ODP & 9.4 & Yes \\
CVE-2026-27577 & 2 & SpreadElement & 9.4 & Yes \\
CVE-2026-27493 & 3 & Form Node double-eval & 9.5 & \textbf{No} \\
\end{longtable}
}

\subsection{CVE-2026-25049 / GHSA-6cqr-8cfr-67f8: Expression Sandbox
Escape (Round
1)}\label{cve-2026-25049-ghsa-6cqr-8cfr-67f8-expression-sandbox-escape-round-1}

\textbf{Package:} \texttt{@n8n/\allowbreak{}tournament}, \texttt{n8n-workflow}\\
\textbf{Affected versions:} All n8n versions prior to 1.123.17 / 2.5.2\\
\textbf{Patched versions:} 1.123.17, 2.5.2

\textbf{CVSS v4.0:} 9.4 ---
\texttt{CVSS:4.\allowbreak{}0/\allowbreak{}AV:N/\allowbreak{}AC:L/\allowbreak{}AT:N/\allowbreak{}PR:L/\allowbreak{}UI:N/\allowbreak{}VC:H/\allowbreak{}VI:H/\allowbreak{}VA:H/\allowbreak{}SC:H/\allowbreak{}SI:H/\allowbreak{}SA:H}

\textbf{Description:} This advisory covers the initial expression
sandbox escape and its immediate bypass. The first exploit used
\texttt{Tagged\allowbreak{}Template\allowbreak{}Expression} as a property accessor,
\texttt{Arrow\allowbreak{}Function\allowbreak{}Expression} as a \texttt{this} carrier, and
\texttt{Error.\allowbreak{}prepare\allowbreak{}Stack\allowbreak{}Trace} for code execution. Each node type
evaded the \texttt{Variable\allowbreak{}Polyfill} rewriter because the specific
combination was not anticipated.

After the initial patch (PR \#23560), a second bypass four days later
used \texttt{Object.\allowbreak{}define\allowbreak{}Property} via \texttt{Call\allowbreak{}Expression} to
manipulate the prototype chain. Different code path, same result: full
RCE. This bypass prompted the comprehensive 9-fix hardening in v2.4.0.

\subsection{CVE-2026-27577 / GHSA-vpcf-gvg4-6qwr: SpreadElement Sandbox
Escape (Round
2)}\label{cve-2026-27577-ghsa-vpcf-gvg4-6qwr-spreadelement-sandbox-escape-round-2}

\textbf{Package:} \texttt{@n8n/\allowbreak{}tournament}\\
\textbf{Component:} \texttt{src/\allowbreak{}Variable\allowbreak{}Polyfill.\allowbreak{}ts},
\texttt{visit\allowbreak{}Identifier} switch statement\\
\textbf{Affected versions:} All n8n versions through 2.10.0 / 2.9.2 /
1.123.21\\
\textbf{Patched versions:} 2.10.1, 2.9.3, 1.123.22

\textbf{CVSS v4.0:} 9.4 ---
\texttt{CVSS:4.\allowbreak{}0/\allowbreak{}AV:N/\allowbreak{}AC:L/\allowbreak{}AT:N/\allowbreak{}PR:L/\allowbreak{}UI:N/\allowbreak{}VC:H/\allowbreak{}VI:H/\allowbreak{}VA:H/\allowbreak{}SC:H/\allowbreak{}SI:H/\allowbreak{}SA:H}

\textbf{Description:} The \texttt{Variable\allowbreak{}Polyfill} transformer in
\texttt{@n8n/\allowbreak{}tournament} rewrites global identifiers into sandboxed
lookups by checking each identifier's parent AST node type against a
switch statement covering 25 node types. \texttt{Spread\allowbreak{}Element} is
absent from the switch. When \texttt{process} appears inside a spread
expression, the parent node type is \texttt{Spread\allowbreak{}Element}, the switch
falls through to \texttt{default} (a no-op \texttt{assert\allowbreak{}Never}), and
the bare \texttt{process} identifier resolves from the real Node.js
global scope. This operates at the compilation stage, before any runtime
sanitizer runs.

\textbf{Payload:} A one-line expression using spread syntax to access
the \texttt{process} global, then calling \texttt{get\allowbreak{}Builtin\allowbreak{}Module} to
obtain \texttt{child\_process} and execute arbitrary commands.

\textbf{Key insight:} This bypass defeated all nine security fixes
shipped in v2.4.0 because it operates at the compilation stage, before
any runtime sanitizer executes.

{\raggedright
\textbf{Why TypeScript missed it:} In \texttt{ast-types},
\texttt{Spread\allowbreak{}Element} is a standalone node type not included in
\texttt{ExpressionKind\ \textbar{}\ StatementKind\ \textbar{}\ PropertyKind\ \textbar{}\ PatternKind\ \textbar{}\ VariableDeclaratorKind\ \textbar{}\ CatchClauseKind}.
TypeScript considers the switch exhaustive; at runtime, unhandled parent
types silently fall through.\par}

\subsection{CVE-2026-27493 / GHSA-75g8-rv7v-32f7: Form Node
Double-Evaluation}\label{cve-2026-27493-ghsa-75g8-rv7v-32f7-form-node-double-evaluation}

\textbf{Package:} \texttt{n8n} (nodes-base)\\
\textbf{Components:}\\
- \texttt{packages/\allowbreak{}nodes-base/\allowbreak{}nodes/\allowbreak{}Form/\allowbreak{}Form.\allowbreak{}node.\allowbreak{}ts} (line 374: first
eval)\\
- \texttt{packages/\allowbreak{}nodes-base/\allowbreak{}nodes/\allowbreak{}Form/\allowbreak{}utils/\allowbreak{}utils.\allowbreak{}ts} (lines 119-139:
second eval via \texttt{prepare\allowbreak{}Form\allowbreak{}Fields})\\
\textbf{Affected versions:} n8n \textgreater= 1.0.0 (when Form nodes
were introduced) through 2.10.0 / 2.9.2 / 1.123.21\\
\textbf{Patched versions:} 2.10.1, 2.9.3, 1.123.22

\textbf{CVSS v4.0:} 9.5 ---
\texttt{CVSS:4.\allowbreak{}0/\allowbreak{}AV:N/\allowbreak{}AC:L/\allowbreak{}AT:P/\allowbreak{}PR:N/\allowbreak{}UI:N/\allowbreak{}VC:H/\allowbreak{}VI:H/\allowbreak{}VA:H/\allowbreak{}SC:H/\allowbreak{}SI:H/\allowbreak{}SA:H}

\textbf{Description:} A second-order expression injection in multi-step
Form nodes. Pass 1 (\texttt{get\allowbreak{}Node\allowbreak{}Parameter}) resolves configured
expressions, injecting attacker-controlled form data into HTML content.
Pass 2 (\texttt{prepare\allowbreak{}Form\allowbreak{}Fields} via \texttt{get\allowbreak{}Resolvables} +
\texttt{evaluate\allowbreak{}Expression}) scans the already-resolved HTML for
expression patterns and evaluates them. Attacker input containing
expression syntax survives Pass 1 as a string and is executed as code in
Pass 2. Form endpoints (\texttt{/\allowbreak{}form/\allowbreak{}*}) are unauthenticated by design.

\textbf{Attack flow:} 1. POST payload to
\texttt{/form/\textless{}path\textgreater{}} (no auth) 2. GET
\texttt{/form-waiting/\textless{}executionId\textgreater{}} 3. Pass 1
injects payload into HTML; Pass 2 evaluates it 4. RCE via sandbox escape
or data exfiltration via expression context

\textbf{Patch:} n8n removed the second expression evaluation pass from
\texttt{prepare\allowbreak{}Form\allowbreak{}Fields} entirely (PR \#26214). Single-pass rendering
eliminates the injection.

\section{Disclosure
Timeline}\label{appendix-b-disclosure-timeline}

All vulnerabilities were reported to n8n's security team through
coordinated disclosure. n8n responded to each report within 24-48 hours.

{\def\LTcaptype{none} 
\begin{longtable}[]{@{}
  >{\raggedright\arraybackslash}p{(\linewidth - 4\tabcolsep) * \real{0.3000}}
  >{\raggedright\arraybackslash}p{(\linewidth - 4\tabcolsep) * \real{0.3500}}
  >{\raggedright\arraybackslash}p{(\linewidth - 4\tabcolsep) * \real{0.3500}}@{}}
\toprule\noalign{}
\begin{minipage}[b]{\linewidth}\raggedright
Date
\end{minipage} & \begin{minipage}[b]{\linewidth}\raggedright
Actor
\end{minipage} & \begin{minipage}[b]{\linewidth}\raggedright
Event
\end{minipage} \\
\midrule\noalign{}
\endhead
\bottomrule\noalign{}
\endlastfoot
December 21, 2025 & Pillar & Round 1 discovered: template literal +
arrow function + \texttt{Error.\allowbreak{}prepare\allowbreak{}Stack\allowbreak{}Trace} chain
(CVE-2026-25049) \\
December 21, 2025 & Pillar & Reported to n8n security team via
responsible disclosure \\
December 22, 2025 & n8n & Confirms vulnerability, acknowledges cloud
impact \\
December 23, 2025 & n8n & Deploys initial patch (PR \#23560), fixing
template literal handling \\
December 24, 2025 & Pillar & Round 1.5: \texttt{Object.\allowbreak{}define\allowbreak{}Property}
bypass discovered (covered by CVE-2026-25049) \\
December 24, 2025 & Pillar & Bypass reported to n8n \\
January 2026 & n8n & Ships v2.4.0 with 9 security fixes:
\texttt{Prototype\allowbreak{}Sanitizer}, \texttt{Function\allowbreak{}This\allowbreak{}Sanitizer}, expanded
blocklists, additional AST node coverage \\
February 3, 2026 & Joint & Coordinated public disclosure of Round 1
findings \\
February 7, 2026 & Pillar & Round 2: SpreadElement bypass discovered
(CVE-2026-27577). Single expression defeats all 9 fixes from v2.4.0.
Operates at compilation stage, before runtime sanitizers execute \\
February 7, 2026 & Pillar & Form Node double-evaluation discovered
(CVE-2026-27493). Unauthenticated RCE via public form endpoint \\
February 7, 2026 & Pillar & Round 2 findings reported to n8n \\
February 25, 2026 & n8n & Publishes patches (2.10.1, 2.9.3, 1.123.22)
and GHSA advisories. Removes second evaluation pass from
\texttt{prepare\allowbreak{}Form\allowbreak{}Fields} (PR \#26214). Adds \texttt{process},
\texttt{require}, \texttt{module}, \texttt{Buffer} to blocked identifier
list \\
\end{longtable}
}

\subsection{Patch Evolution}\label{patch-evolution}

The patches evolved in scope and approach across rounds, reflecting
genuine learning:

\begin{itemize}
\tightlist
\item
  \textbf{Round 1 patch (PR \#23560):} Targeted fix for the specific
  template literal + \texttt{Error.\allowbreak{}prepare\allowbreak{}Stack\allowbreak{}Trace} vector. Correct
  and fast (48 hours, over holidays).
\item
  \textbf{v2.4.0 hardening (January):} Nine security fixes across
  multiple code paths --- \texttt{Prototype\allowbreak{}Sanitizer},
  \texttt{Function\allowbreak{}This\allowbreak{}Sanitizer}, expanded blocklists, additional AST
  node coverage, immutable \texttt{\_\_sanitize}, destructuring
  protection. A systematic audit across multiple code paths.
\item
  \textbf{Round 2 patch (PR \#26214):} Added \texttt{Spread\allowbreak{}Element} to
  the switch statement AND blocked \texttt{process}, \texttt{require},
  \texttt{module}, \texttt{Buffer} as identifiers. Also removed the
  second evaluation pass from \texttt{prepare\allowbreak{}Form\allowbreak{}Fields} entirely for
  the Form Node fix --- eliminating the double-evaluation rather than
  trying to sanitize it.
\end{itemize}

Each iteration was more comprehensive than the last. The structural
problem is that even thorough enumeration cannot anticipate the next
unhandled node type.

\subsection{Key Observations}\label{key-observations}

\begin{enumerate}
\def\labelenumi{\arabic{enumi}.}
\item
  \textbf{Response time:} n8n acknowledged every report within 24-48
  hours, including over the December holidays. Patches were deployed
  within days of confirmation.
\item
  \textbf{Fix quality:} Each individual fix was correct for the specific
  bypass it addressed. The v2.4.0 release was a thorough hardening
  effort across multiple code paths. The final patch (PR \#26214) took
  the strongest approach available: blocking known-dangerous globals at
  the identifier level and eliminating double-evaluation entirely.
\item
  \textbf{Structural pattern:} Despite correct fixes, each subsequent
  bypass was structurally inevitable. Round 2's SpreadElement bypass
  operated at a different layer (compilation vs.~runtime) than all 9
  fixes from v2.4.0. The fix-bypass cycle is a property of the AST
  enumeration approach, not of n8n's engineering quality.
\item
  \textbf{Independent vulnerabilities:} The Form Node double-evaluation
  (CVE-2026-27493) is architecturally independent from the sandbox
  escapes. Patching the sandbox does not fix the injection. The
  double-evaluation is a logic bug in the Form node's rendering
  pipeline.
\end{enumerate}

\section{AST Coverage Analysis
Methodology}\label{appendix-c-ast-coverage-analysis-methodology}

This appendix describes the five-step methodology for quantifying the
attack surface of an expression engine's AST-based sandbox. The
methodology was developed during our analysis of n8n's
\texttt{@n8n/\allowbreak{}tournament} package and generalizes to any expression
engine that constrains execution through AST node type enumeration.

\subsection{Step 1: Extract the AST Rewriting
Logic}\label{step-1-extract-the-ast-rewriting-logic}

Identify the expression engine's compilation or transformation pipeline.
Locate the component that rewrites or restricts AST nodes before
execution.

\textbf{n8n example:} The \texttt{@n8n/\allowbreak{}tournament} package compiles
expressions through a \texttt{Variable\allowbreak{}Polyfill} transformer. The
\texttt{visit\allowbreak{}Identifier} method checks each identifier's parent AST node
type against a switch statement to determine whether and how to rewrite
it.

\textbf{What to look for:} - AST visitors or transformers that modify
identifier resolution - Switch statements, if-else chains, or lookup
tables keyed on node types - Default/fallback behavior for unrecognized
node types

\subsection{Step 2: Enumerate Handled Node
Types}\label{step-2-enumerate-handled-node-types}

Extract the complete list of AST node types the rewriter explicitly
handles.

\textbf{n8n example:} The \texttt{visit\allowbreak{}Identifier} switch statement
covers 25 node types:

\begin{Verbatim}[breaklines=true,fontsize=\small]
MemberExpression, CallExpression, BinaryExpression,
LogicalExpression, ConditionalExpression, UnaryExpression,
VariableDeclarator, AssignmentExpression, Property,
ArrayExpression, SequenceExpression, TemplateLiteral,
TaggedTemplateExpression, ReturnStatement, IfStatement,
SwitchCase, ThrowStatement, ForStatement, ForInStatement,
ForOfStatement, WhileStatement, DoWhileStatement,
ExpressionStatement, NewExpression, UpdateExpression
\end{Verbatim}

\subsection{Step 3: Enumerate the Full Language Specification Node
Types}\label{step-3-enumerate-the-full-language-specification-node-types}

Map the complete set of AST node types defined in the target language
specification.

\textbf{For ECMAScript / JavaScript:} - Primary source: ESTree
specification \cite{ref16} - Secondary source: Babel AST documentation -
Tertiary source: \texttt{ast-types} package type definitions \cite{ref17}

The ESTree specification defines 72 distinct AST node types. Key
categories:

{\def\LTcaptype{none} 
\begin{longtable}[]{@{}
  >{\raggedright\arraybackslash}p{(\linewidth - 2\tabcolsep) * \real{0.3448}}
  >{\raggedright\arraybackslash}p{(\linewidth - 2\tabcolsep) * \real{0.6552}}@{}}
\toprule\noalign{}
\begin{minipage}[b]{\linewidth}\raggedright
Category
\end{minipage} & \begin{minipage}[b]{\linewidth}\raggedright
Example Node Types
\end{minipage} \\
\midrule\noalign{}
\endhead
\bottomrule\noalign{}
\endlastfoot
Expressions & \texttt{Spread\allowbreak{}Element}, \texttt{Yield\allowbreak{}Expression},
\texttt{Await\allowbreak{}Expression}, \texttt{Import\allowbreak{}Expression},
\texttt{Chain\allowbreak{}Expression}, \texttt{Class\allowbreak{}Expression} \\
Declarations & \texttt{Class\allowbreak{}Declaration},
\texttt{Export\allowbreak{}Named\allowbreak{}Declaration}, \texttt{Import\allowbreak{}Declaration} \\
Patterns & \texttt{Rest\allowbreak{}Element}, \texttt{Assignment\allowbreak{}Pattern},
\texttt{Object\allowbreak{}Pattern}, \texttt{Array\allowbreak{}Pattern} \\
Statements & \texttt{Labeled\allowbreak{}Statement}, \texttt{With\allowbreak{}Statement},
\texttt{Try\allowbreak{}Statement} \\
Clauses & \texttt{Catch\allowbreak{}Clause}, \texttt{Switch\allowbreak{}Case} \\
Miscellaneous & \texttt{Spread\allowbreak{}Property}, \texttt{Method\allowbreak{}Definition},
\texttt{Meta\allowbreak{}Property} \\
\end{longtable}
}

\subsection{Step 4: Classify Unhandled Types by Exploitation
Potential}\label{step-4-classify-unhandled-types-by-exploitation-potential}

For each unhandled node type, assess whether it can:

\begin{enumerate}
\def\labelenumi{\arabic{enumi}.}
\tightlist
\item
  \textbf{Carry a global reference:} Can this node type contain an
  identifier that would resolve from the global scope if not rewritten?
  (e.g., \texttt{Spread\allowbreak{}Element} containing \texttt{process})
\item
  \textbf{Bypass identifier rewriting:} Does this node type create a
  context where the rewriter's assumptions about parent-child
  relationships don't hold?
\item
  \textbf{Enable code execution:} Can this node type directly or
  indirectly invoke functions, access prototypes, or reach restricted
  APIs?
\end{enumerate}

\textbf{Severity classification:}

{\def\LTcaptype{none} 
\begin{longtable}[]{@{}
  >{\raggedright\arraybackslash}p{(\linewidth - 2\tabcolsep) * \real{0.5000}}
  >{\raggedright\arraybackslash}p{(\linewidth - 2\tabcolsep) * \real{0.5000}}@{}}
\toprule\noalign{}
\begin{minipage}[b]{\linewidth}\raggedright
Severity
\end{minipage} & \begin{minipage}[b]{\linewidth}\raggedright
Criteria
\end{minipage} \\
\midrule\noalign{}
\endhead
\bottomrule\noalign{}
\endlastfoot
Critical & Node type can carry a bare global reference that escapes
rewriting. Direct path to RCE. \\
High & Node type can alter control flow or scope in ways that bypass
runtime sanitizers. \\
Medium & Node type can access properties or invoke behavior not covered
by blocklists. \\
Low & Node type is syntactically valid but unlikely to carry exploitable
semantics in the sandbox context. \\
\end{longtable}
}

\textbf{n8n example:} \texttt{Spread\allowbreak{}Element} is classified as
\textbf{Critical} -- it carries an identifier (\texttt{process}) whose
parent type is not in the switch statement, causing the identifier to
pass through unrewritten. This was confirmed by CVE-2026-27577.

\subsection{Step 5: Score the Coverage
Gap}\label{step-5-score-the-coverage-gap}

Produce a coverage report:

\begin{Shaded}
\begin{Highlighting}[]
\NormalTok{Coverage Report: @n8n/tournament VariablePolyfill}
\NormalTok{=================================================}
\NormalTok{Total ESTree node types:         72}
\NormalTok{Handled by visitIdentifier:      25}
\NormalTok{Unhandled:                       47}
\NormalTok{  {-} Critical severity:             2  (SpreadElement, RestElement)}
\NormalTok{  {-} High severity:                 4  (YieldExpression, AwaitExpression, ...)}
\NormalTok{  {-} Medium severity:              10}
\NormalTok{  {-} Low severity:                 31}

\NormalTok{Coverage ratio: 35\% (25/72)}
\NormalTok{Critical gap: SpreadElement {-}{-} CONFIRMED BYPASS (CVE{-}2026{-}27577)}
\end{Highlighting}
\end{Shaded}

The coverage ratio is the primary metric. A ratio below 50\% indicates
that the sandbox's enumeration covers less than half the language's
instruction set -- the weird machine has more undefended instructions
than defended ones. For n8n, the 35\% coverage ratio means nearly
two-thirds of the language's node types are undefended.

\subsection{Generalization to Other
Platforms}\label{generalization-to-other-platforms}

The methodology applies to any expression engine that constrains
execution through AST node type enumeration:

{\def\LTcaptype{none} 
\begin{longtable}[]{@{}
  >{\raggedright\arraybackslash}p{(\linewidth - 6\tabcolsep) * \real{0.1667}}
  >{\raggedright\arraybackslash}p{(\linewidth - 6\tabcolsep) * \real{0.3000}}
  >{\raggedright\arraybackslash}p{(\linewidth - 6\tabcolsep) * \real{0.3167}}
  >{\raggedright\arraybackslash}p{(\linewidth - 6\tabcolsep) * \real{0.2167}}@{}}
\toprule\noalign{}
\begin{minipage}[b]{\linewidth}\raggedright
Platform
\end{minipage} & \begin{minipage}[b]{\linewidth}\raggedright
Expression Engine
\end{minipage} & \begin{minipage}[b]{\linewidth}\raggedright
Sandboxing Approach
\end{minipage} & \begin{minipage}[b]{\linewidth}\raggedright
Applicable?
\end{minipage} \\
\midrule\noalign{}
\endhead
\bottomrule\noalign{}
\endlastfoot
n8n & @n8n/tournament & AST rewriting (VariablePolyfill) & Yes --
reference implementation \\
Apache Airflow & Jinja2 & Template sandbox (SandboxedEnvironment) & Yes
-- enumerate allowed filters/functions \\
GitHub Actions & Shell expansion & No sandbox (\$\{\{ \}\} evaluated in
shell) & Partially -- no AST layer to analyze \\
Custom engines & Various & Various & Yes, if AST-based restriction is
the security boundary \\
\end{longtable}
}

The tool to be released alongside this paper automates Steps 1-5 for
\texttt{@n8n/\allowbreak{}tournament} as the reference implementation and provides a
pluggable parser interface for other expression engine architectures.

\bibliographystyle{unsrtnat}
\bibliography{references}
\end{document}

%% file: preamble.tex
\usepackage{tikz}
\usetikzlibrary{arrows.meta,positioning,fit,backgrounds,calc,shapes.geometric,
                decorations.pathreplacing}

\newcommand{\wmsteptitle}[1]{{\sffamily\small\bfseries #1}}

\newcommand{\wmiconno}[1]{}
\IfFileExists{fontawesome5.sty}{%
  \usepackage{fontawesome5}%
  \let\wmicon\wmiconyes
}{%
  \let\wmicon\wmiconno
}

\definecolor{wmred}{HTML}{C53030}
\definecolor{wmredfill}{HTML}{FED7D7}
\definecolor{wmredbg}{HTML}{FFF5F5}
\definecolor{wmredsolid}{HTML}{E53E3E}
\definecolor{wmamber}{HTML}{B7791F}
\definecolor{wmamberfill}{HTML}{FEFCBF}
\definecolor{wmamberbg}{HTML}{FFFBEB}
\definecolor{wmorange}{HTML}{C05621}
\definecolor{wmorangefill}{HTML}{FEEBC8}
\definecolor{wmgreen}{HTML}{276749}
\definecolor{wmgreenfill}{HTML}{C6F6D5}
\definecolor{wmgreenbg}{HTML}{F0FFF4}
\definecolor{wmblue}{HTML}{2B6CB0}
\definecolor{wmbluefill}{HTML}{BEE3F8}
\definecolor{wmbluebg}{HTML}{EBF8FF}
\definecolor{wmpurple}{HTML}{6B46C1}
\definecolor{wmpurplefill}{HTML}{E9D8FD}
\definecolor{wmpurplebg}{HTML}{FAF5FF}
\definecolor{wmslate}{HTML}{1A202C}
\definecolor{wmslatefill}{HTML}{2D3748}
\definecolor{wmgrey}{HTML}{4A5568}
\definecolor{wmgreylight}{HTML}{A0AEC0}
\definecolor{wmgreybg}{HTML}{EDF2F7}
\definecolor{wmredsoft}{HTML}{FC8181}
\definecolor{wmorangesoft}{HTML}{F6AD55}
\definecolor{wmyellow}{HTML}{ECC94B}
\definecolor{wmgold}{HTML}{D69E2E}
\definecolor{wmgreensoft}{HTML}{48BB78}

\tikzset{
  wmbase/.style={
    draw, rounded corners=7pt, align=center, font=\sffamily\footnotesize,
    inner sep=6pt, line width=0.8pt,
    execute at begin node={\hyphenpenalty=10000\exhyphenpenalty=10000\relax}},
  wmstage/.style={wmbase, font=\sffamily\small\bfseries, minimum height=1.0cm},
  wmatk/.style   ={wmstage, draw=wmred,    fill=wmredbg},
  wmwarn/.style  ={wmstage, draw=wmamber,  fill=wmamberbg},
  wmsafe/.style  ={wmstage, draw=wmgreen,  fill=wmgreenbg},
  wminfo/.style  ={wmstage, draw=wmblue,   fill=wmbluebg},
  wmint/.style   ={wmstage, draw=wmpurple, fill=wmpurplebg},
  wmdark/.style  ={wmstage, draw=wmslate,  fill=wmslate, text=white},
  wmdetail/.style       ={wmbase, rounded corners=4pt, font=\sffamily\scriptsize},
  wmdetail atk/.style   ={wmdetail, draw=wmred,    fill=wmredfill},
  wmdetail warn/.style  ={wmdetail, draw=wmamber,  fill=wmamberfill},
  wmdetail safe/.style  ={wmdetail, draw=wmgreen,  fill=wmgreenfill},
  wmdetail info/.style  ={wmdetail, draw=wmblue,   fill=wmbluefill},
  wmdetail int/.style   ={wmdetail, draw=wmpurple, fill=wmpurplefill},
  wmdetail dark/.style  ={wmdetail, draw=wmslate,  fill=wmslatefill, text=wmgreybg},
  wmdetail orange/.style={wmdetail, draw=wmorange, fill=wmorangefill},
  wmbadge/.style        ={wmbase, rounded corners=4pt,
                          font=\sffamily\scriptsize\bfseries, text=white},
  wmgroup/.style={draw, rounded corners=9pt, line width=1.0pt, inner sep=9pt},
  wmgrouptitle/.style={font=\sffamily\small\bfseries},
  wmnote/.style={wmbase, rounded corners=5pt, draw=wmgrey, fill=wmgreybg,
                 text=wmslatefill, font=\sffamily\scriptsize\itshape},
  wmcallout/.style={wmbase, rounded corners=5pt, densely dashed,
                    draw=wmred, fill=wmredbg, text=wmred,
                    font=\sffamily\scriptsize\bfseries},
  wmflow/.style={-{Stealth[length=2.6mm]}, line width=1.1pt, draw=wmgrey},
  wmflow atk/.style={wmflow, draw=wmred},
  wmflow warn/.style={wmflow, draw=wmamber},
  wmflow safe/.style={wmflow, draw=wmgreen},
  wmflow int/.style={wmflow, draw=wmpurple},
  wmflow dark/.style={wmflow, draw=wmslate},
  wmghost/.style={-{Stealth[length=2.2mm]}, line width=0.7pt,
                  draw=wmgreylight, densely dashed},
  wmhint/.style={-{Stealth[length=2.2mm]}, line width=0.9pt,
                 draw=wmred, densely dashed},
  wmedgelabel/.style={font=\sffamily\scriptsize, text=wmgrey,
                      align=center, inner sep=2pt},
}

%% file: trust-boundaries.tex
\begin{figure}
\centering
\begin{adjustbox}{max width=\linewidth,max totalheight=0.78\textheight,center}
\begin{tikzpicture}[
    inner/.style={wmdetail, minimum width=3.5cm, minimum height=1.0cm},
  ]

  \node[inner, draw=wmamber, fill=wmamberfill]
    (expr) at (6.0,1.05) {\wmicon{\faCode}Expression engine};
  \node[inner, draw=wmamber, fill=wmamberfill, below=3mm of expr]
    (ai) {\wmicon{\faRobot}AI workflow logic};
  \node[inner, draw=wmamber, fill=wmamberfill, below=3mm of ai]
    (exec) {\wmicon{\faPlay}Node execution};
  \node[inner, draw=wmorange, fill=wmorangefill, below=3mm of exec,
        font=\sffamily\scriptsize\bfseries]
    (creds) {\wmicon{\faKey}Credential vault};

  \node[wmdetail atk, minimum width=2.9cm, minimum height=1.6cm]
    (users) at (0,-0.9) {\wmicon{\faUsers}\textbf{Untrusted users}\\+ tenants};

  \node[wmdetail info, minimum width=2.9cm, minimum height=2.4cm]
    (targets) at (12.0,-0.9)
    {\wmicon{\faCloud}Cloud APIs\\Databases\\AI providers\\SaaS apps};

  \begin{scope}[on background layer]
    \node[wmgroup, draw=wmred, fill=wmredbg, fit=(users)] (zint) {};
    \node[wmgroup, draw=wmamber, fill=wmamberbg,
          fit=(expr)(ai)(exec)(creds)] (zplat) {};
    \node[wmgroup, draw=wmblue, fill=wmbluebg, fit=(targets)] (zsys) {};
  \end{scope}

  \node[wmgrouptitle, text=wmred,  above=1.5mm of zint.north] {INTERNET};
  \node[wmgrouptitle, text=wmblue, above=1.5mm of zsys.north] {CONNECTED SYSTEMS};
  \node[wmgrouptitle, text=wmamber, above=1.5mm of zplat.north, align=center]
    {PLATFORM PROCESS\\[-2pt]
     {\sffamily\scriptsize\mdseries single process, one trust domain}};

  \draw[wmflow atk] (users.east) -- ++(0.55,0) |- (expr.west);
  \node[wmedgelabel, text=wmred, anchor=north] at (2.62,0.92)
    {public endpoints\\+ auth};
  \draw[wmflow, draw=wmblue] (creds.east) -- ++(0.9,0) |- (targets.west);
  \node[wmedgelabel, text=wmblue, anchor=south] at (9.65,-0.78) {credentials};

  \node[wmnote, below=6mm of zplat.south, text width=8.8cm]
    {A sandbox escape in the expression engine is already inside the
     credential vault's trust domain: no boundary separates them.};

\end{tikzpicture}
\end{adjustbox}
\caption{Trust boundaries in a typical orchestration platform
deployment. The expression engine, credential vault, and workflow
execution share a single process. Sandbox escape equals full credential
access.}
\label{fig:trust-boundaries}
\end{figure}
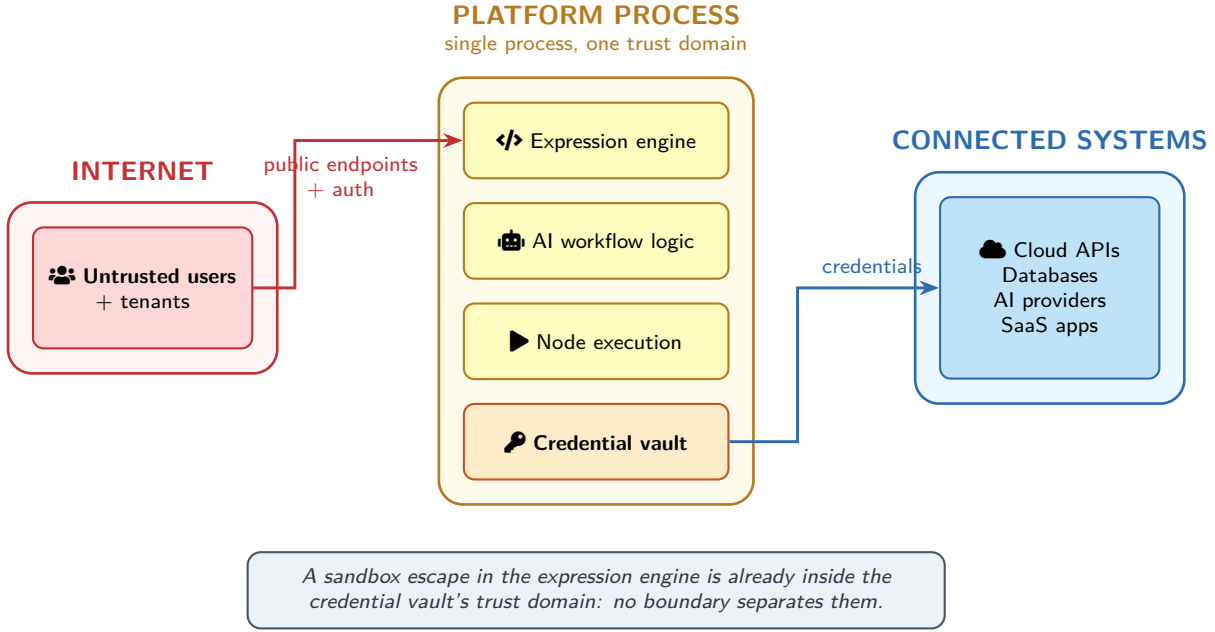

%% file: compilation-pipeline.tex
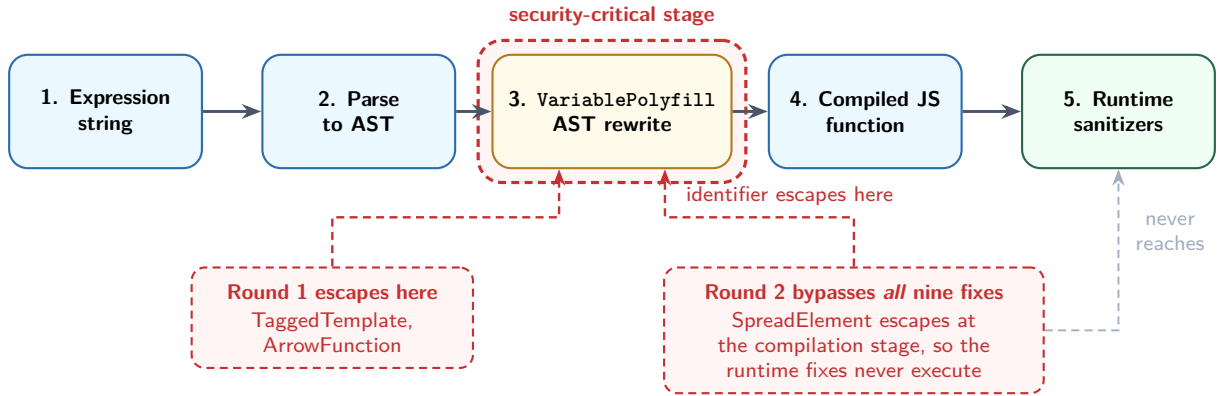
\begin{figure}
\centering
\begin{adjustbox}{max width=\linewidth,max totalheight=0.78\textheight,center}
\begin{tikzpicture}[
    stage/.style={wmstage, minimum width=2.55cm, minimum height=1.5cm,
                  font=\sffamily\scriptsize\bfseries},
  ]

  \node[stage, draw=wmblue,  fill=wmbluebg]  (s1) at ( 0.0,0) {1. Expression\\string};
  \node[stage, draw=wmblue,  fill=wmbluebg]  (s2) at ( 3.35,0) {2. Parse\\to AST};
  \node[stage, draw=wmamber, fill=wmamberbg] (s3) at ( 6.70,0)
    {3. \texttt{VariablePolyfill}\\AST rewrite};
  \node[stage, draw=wmblue,  fill=wmbluebg]  (s4) at (10.05,0) {4. Compiled JS\\function};
  \node[stage, draw=wmgreen, fill=wmgreenbg] (s5) at (13.40,0) {5. Runtime\\sanitizers};

  \foreach \a/\b in {s1/s2,s2/s3,s3/s4,s4/s5}
    \draw[wmflow] (\a) -- (\b);

  \begin{scope}[on background layer]
    \node[wmgroup, draw=wmred, fill=wmredbg, densely dashed, line width=1.2pt,
          fit=(s3), inner sep=5pt] (gap) {};
  \end{scope}
  \node[wmedgelabel, text=wmred, font=\sffamily\scriptsize\bfseries,
        above=1mm of gap.north] {security-critical stage};

  \node[wmcallout, text width=3.3cm, anchor=north] (r1) at (3.0,-2.05)
    {Round 1 escapes here\\[1pt]
     {\sffamily\scriptsize\mdseries TaggedTemplate, ArrowFunction}};
  \node[wmcallout, text width=4.6cm, anchor=north] (r2) at (9.9,-2.05)
    {Round 2 bypasses \emph{all} nine fixes\\[1pt]
     {\sffamily\scriptsize\mdseries SpreadElement escapes at the compilation
      stage, so the runtime fixes never execute}};

  \draw[wmhint] (r1.north) -- ++(0,0.65) -| ([xshift=-7mm]s3.south);
  \draw[wmhint] (r2.north) -- ++(0,0.65) -| ([xshift=7mm]s3.south);
  \node[wmedgelabel, text=wmred, anchor=south] at (9.05,-1.32)
    {identifier escapes here};

  \draw[wmghost] (r2.east) -| (s5.south);
  \node[wmedgelabel, text=wmgreylight, anchor=west] at (13.55,-1.6)
    {never\\reaches};

\end{tikzpicture}
\end{adjustbox}
\caption{Expression sandbox compilation pipeline. Round 1 escapes
exploit the AST rewriting stage; Round 2's SpreadElement bypass escapes
at the same stage, so all nine runtime sanitizer fixes never execute.}
\label{fig:compilation-pipeline}
\end{figure}

%% file: fix-bypass-cycle.tex
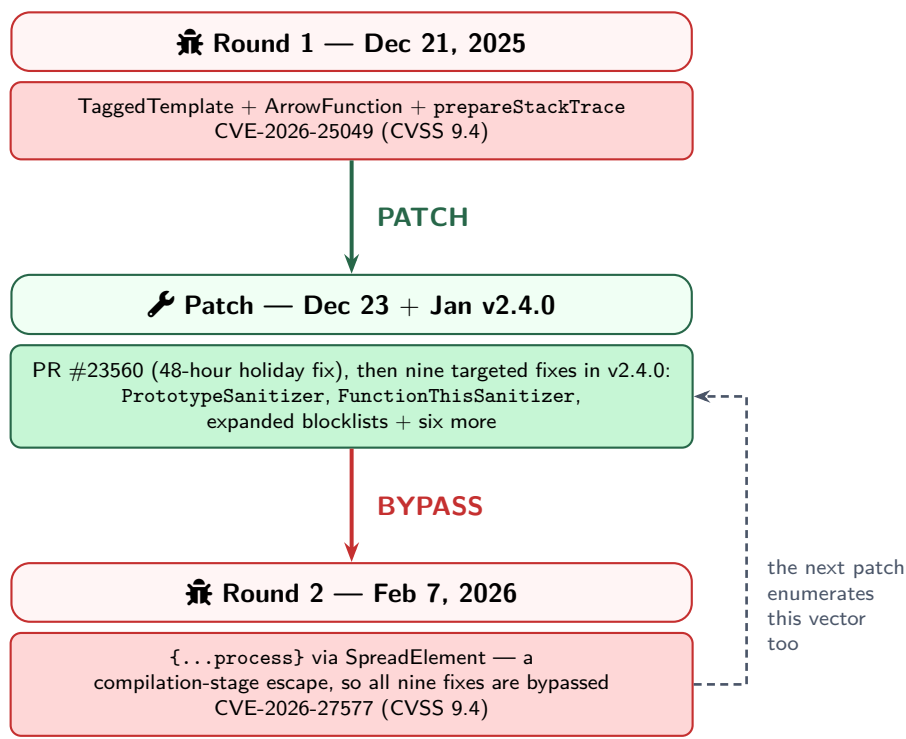
\begin{figure}
\centering
\begin{adjustbox}{max width=\linewidth,max totalheight=0.78\textheight,center}
\begin{tikzpicture}[
    head/.style={wmstage, minimum width=9.0cm, minimum height=0.75cm,
                 font=\sffamily\small\bfseries},
    body/.style={wmdetail, text width=8.6cm, align=center},
    wmstep/.style={wmedgelabel, font=\sffamily\small\bfseries, inner sep=3pt},
  ]

  \node[head, draw=wmred, fill=wmredbg] (h1) at (0,0)
    {\wmicon{\faBug}Round 1 --- Dec 21, 2025};
  \node[body, draw=wmred, fill=wmredfill, below=1.2mm of h1] (b1)
    {TaggedTemplate $+$ ArrowFunction $+$ \texttt{prepareStackTrace}\\
     CVE-2026-25049 (CVSS 9.4)};

  \node[head, draw=wmgreen, fill=wmgreenbg, below=1.5cm of b1] (h2)
    {\wmicon{\faWrench}Patch --- Dec 23 $+$ Jan v2.4.0};
  \node[body, draw=wmgreen, fill=wmgreenfill, below=1.2mm of h2] (b2)
    {PR \#23560 (48-hour holiday fix), then nine targeted fixes in
     v2.4.0:\\ \texttt{PrototypeSanitizer}, \texttt{FunctionThisSanitizer},
     expanded blocklists $+$ six more};

  \node[head, draw=wmred, fill=wmredbg, below=1.5cm of b2] (h3)
    {\wmicon{\faBug}Round 2 --- Feb 7, 2026};
  \node[body, draw=wmred, fill=wmredfill, below=1.2mm of h3] (b3)
    {\texttt{\{...process\}} via SpreadElement --- a compilation-stage escape,
     so all nine fixes are bypassed\\ CVE-2026-27577 (CVSS 9.4)};

  \draw[wmflow safe, line width=1.5pt] (b1.south) -- (h2.north)
    node[wmstep, text=wmgreen, midway, right=2mm] {PATCH};
  \draw[wmflow atk, line width=1.5pt] (b2.south) -- (h3.north)
    node[wmstep, text=wmred, midway, right=2mm] {BYPASS};

  \draw[wmhint, draw=wmgrey, line width=1.0pt]
    (b3.east) -- ++(0.7,0) |- (b2.east);
  \node[wmedgelabel, anchor=west, align=left, text=wmgrey]
    at ($(b3.east)+(0.9,1.05)$)
    {the next patch\\enumerates\\this vector\\too};

\end{tikzpicture}
\end{adjustbox}
\caption{Fix-bypass escalation cycle. Each patch was correct for the
known vector; each was rendered irrelevant by the next bypass using a
different weird machine instruction.}
\label{fig:fix-bypass-cycle}
\end{figure}

%% file: trust-laundering-pipeline.tex
\begin{figure}
\centering
\begin{adjustbox}{max width=\linewidth,max totalheight=0.78\textheight,center}
\begin{tikzpicture}[
    head/.style={wmstage, minimum width=3.6cm, minimum height=0.95cm,
                 font=\sffamily\scriptsize\bfseries},
    wm/.style={wmdetail, text width=3.25cm, align=center, minimum height=1.35cm},
    badge/.style={wmbadge, text width=3.25cm, align=center, minimum height=0.85cm},
  ]

  \node[head, draw=wmred, fill=wmredbg] (n1) at (0,0)
    {\wmicon{\faWpforms}Form Trigger\\\texttt{/form/support}};
  \node[wm, draw=wmred, fill=wmredfill, below=1.5mm of n1] (w1)
    {Public endpoint\\No authentication\\by design};
  \node[badge, draw=wmred, fill=wmredsoft, below=1.5mm of w1] (t1)
    {UNTRUSTED\\user input};

  \node[head, draw=wmamber, fill=wmamberbg] (n2) at (4.25,0)
    {\wmicon{\faCode}Code Node\\(expression engine)};
  \node[wm, draw=wmamber, fill=wmamberfill, below=1.5mm of n2] (w2)
    {\textbf{WM1: ECMAScript}\\Deterministic\\Full attacker control};
  \node[badge, draw=wmorange, fill=wmorangesoft, below=1.5mm of w2] (t2)
    {Taint stripped:\\``workflow data''};

  \node[head, draw=wmgold, fill=wmamberfill] (n3) at (8.5,0)
    {\wmicon{\faComments}OpenAI Chat\\(LLM node)};
  \node[wm, draw=wmgold, fill=wmamberfill, below=1.5mm of n3] (w3)
    {\textbf{WM2: natural language}\\Non-deterministic\\Partial attacker control};
  \node[badge, draw=wmamber, fill=wmyellow, text=wmslate, below=1.5mm of w3] (t3)
    {Taint stripped:\\``trusted context''};

  \node[head, draw=wmgreen, fill=wmgreenbg] (n4) at (12.75,0)
    {\wmicon{\faEnvelope}Gmail Node\\(credentialed action)};
  \node[wm, draw=wmgreen, fill=wmgreenfill, below=1.5mm of n4] (w4)
    {\textbf{WM3: credentialed API}\\Stored OAuth token\\Full platform trust};
  \node[badge, draw=wmgreen, fill=wmgreensoft, below=1.5mm of w4] (t4)
    {FULLY TRUSTED\\API call};

  \foreach \a/\b/\c in {n1/n2/wmorange, n2/n3/wmamber, n3/n4/wmgreen}
    \draw[wmflow, draw=\c, line width=1.5pt] (\a) -- (\b);
  \foreach \x/\c in {2.125/wmorange, 6.375/wmamber, 10.625/wmgreen}
    \node[wmedgelabel, text=\c, anchor=south] at (\x,0.62) {taint stripped};

  \node[wmnote, below=5mm of $(t1.south)!0.5!(t4.south)$, text width=14.6cm]
    {Each transformation strips taint, and no explicit trust decision occurs at
     any boundary. By the final node, attacker-controlled input drives fully
     credentialed actions.};

\end{tikzpicture}
\end{adjustbox}
\caption{Trust laundering pipeline. Attacker-controlled input flows
through four nodes, each transformation stripping taint until the input
drives fully credentialed API calls. No explicit trust decision occurs
at any boundary.}
\label{fig:trust-laundering-pipeline}
\end{figure}
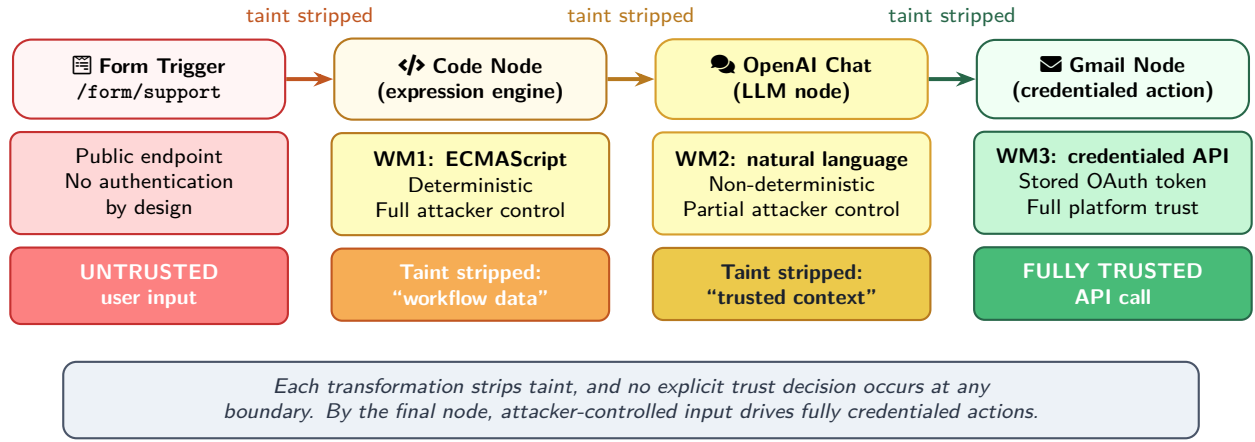

%% file: kill-chain.tex
\begin{figure}
\centering
\begin{adjustbox}{max width=\linewidth,max totalheight=0.78\textheight,center}
\begin{tikzpicture}[
    kstep/.style={wmbase, minimum width=7.0cm, minimum height=1.55cm,
                 text width=6.7cm, align=center, font=\sffamily\scriptsize},
  ]

  \node[kstep, draw=wmred, fill=wmredbg] (k1) at (0,0)
    {\wmsteptitle{\wmicon{\faSyringe}1. Expression injection}\\[2pt]
     CVE-2026-27577 or CVE-2026-27493\\
     SpreadElement / Form double-eval};
  \node[kstep, draw=wmred, fill=wmredbg, below=7mm of k1] (k2)
    {\wmsteptitle{\wmicon{\faTerminal}2. RCE on the n8n process}\\[2pt]
     Sandbox escape yields full\\ Node.js process access};
  \node[kstep, draw=wmamber, fill=wmamberbg, below=7mm of k2] (k3)
    {\wmsteptitle{\wmicon{\faKey}3. Read \texttt{N8N\_ENCRYPTION\_KEY}}\\[2pt]
     \texttt{process.env} access; a single key\\
     encrypts \emph{all} stored credentials};

  \node[kstep, draw=wmamber, fill=wmamberbg] (k4) at (8.4,0)
    {\wmsteptitle{\wmicon{\faLockOpen}4. Decrypt and modify credentials}\\[2pt]
     AES-256-CBC with a deprecated KDF;\\
     swap the AI provider \texttt{baseUrl}
     to an attacker proxy};
  \node[kstep, draw=wmpurple, fill=wmpurplebg, below=7mm of k4] (k5)
    {\wmsteptitle{\wmicon{\faNetworkWired}5. Proxy intercepts every LLM call}\\[2pt]
     Transparent reverse proxy captures API keys
     and payloads, forwards to the real
     provider; no certificate pinning};
  \node[kstep, draw=wmslate, fill=wmslate, text=white, below=7mm of k5] (k6)
    {\wmsteptitle{\wmicon{\faEyeSlash}6. Persistent silent compromise}\\[2pt]
     Survives restarts (DB-persisted);
     all guardrails report green;
     no trace in the n8n UI};

  \draw[wmflow atk,    line width=1.5pt] (k1) -- (k2);
  \draw[wmflow warn,   line width=1.5pt] (k2) -- (k3);
  \draw[wmflow int,    line width=1.5pt] (k4) -- (k5);
  \draw[wmflow dark,   line width=1.5pt] (k5) -- (k6);
  \draw[wmflow warn, line width=1.5pt]
    (k3.east) -- ++(0.55,0) |- (k4.west);

  \node[wmedgelabel, text=wmred, anchor=north, align=center]
    at ($(k3.south)+(0,-0.35)$) {\textbf{escape} $\rightarrow$ \textbf{key}};
  \node[wmedgelabel, text=wmslate, anchor=north, align=center]
    at ($(k6.south)+(0,-0.35)$)
    {\textbf{tamper} $\rightarrow$ \textbf{intercept} $\rightarrow$
     \textbf{persist}};

\end{tikzpicture}
\end{adjustbox}
\caption{AI infrastructure kill-chain. Expression injection escalates
through credential extraction and base URL tampering to persistent,
silent interception of all LLM API calls.}
\label{fig:kill-chain}
\end{figure}
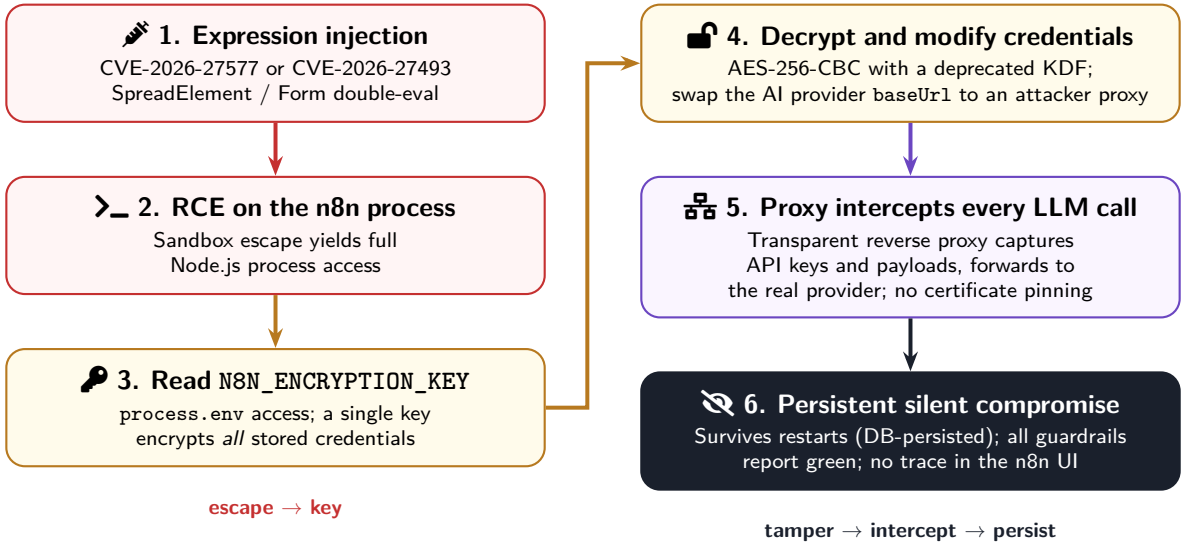

%% file: ast-coverage-gap.tex
%
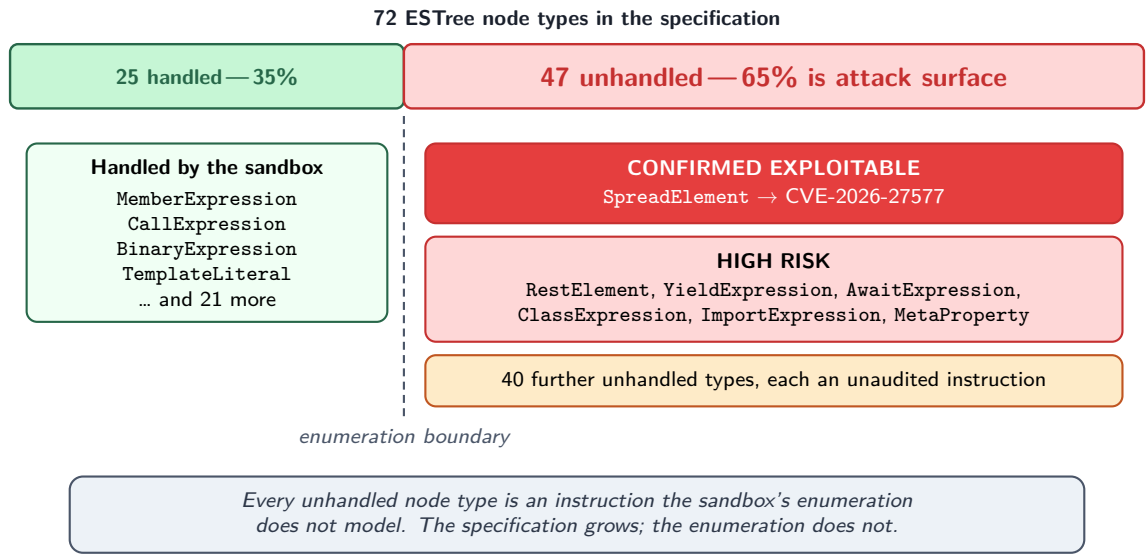
\begin{figure}
\centering
\begin{adjustbox}{max width=\linewidth,max totalheight=0.78\textheight,center}
\begin{tikzpicture}[
    detail/.style={wmdetail, align=center, font=\sffamily\scriptsize},
  ]
  \def\barw{15.0}      
  \def\barsplit{5.21}  
  \def\barh{0.85}      

  \draw[fill=wmgreenfill, draw=wmgreen, line width=0.9pt,
        rounded corners=3pt] (0,0) rectangle (\barsplit,\barh);
  \draw[fill=wmredfill, draw=wmred, line width=0.9pt,
        rounded corners=3pt] (\barsplit,0) rectangle (\barw,\barh);

  \node[font=\sffamily\scriptsize\bfseries, text=wmgreen]
    at (\barsplit/2,\barh/2) {25 handled\,---\,35\%};
  \node[font=\sffamily\small\bfseries, text=wmred]
    at ({(\barsplit+\barw)/2},\barh/2)
    {47 unhandled\,---\,65\% is attack surface};

  \node[wmedgelabel, anchor=south, font=\sffamily\scriptsize\bfseries,
        text=wmslate] at (\barw/2,\barh+0.12)
    {72 ESTree node types in the specification};

  \node[detail, draw=wmgreen, fill=wmgreenbg, text width=4.35cm,
        anchor=north] (handled) at (\barsplit/2,-0.45)
    {\textbf{Handled by the sandbox}\\[2pt]
     \texttt{MemberExpression}\\ \texttt{CallExpression}\\
     \texttt{BinaryExpression}\\ \texttt{TemplateLiteral}\\
     \dots\ and 21 more};

  \node[detail, draw=wmred, fill=wmredsolid, text=white, text width=8.8cm,
        anchor=north, font=\sffamily\scriptsize\bfseries]
        (exploited) at ({(\barsplit+\barw)/2},-0.45)
    {CONFIRMED EXPLOITABLE\\[1pt]
     {\sffamily\scriptsize\mdseries \texttt{SpreadElement}
      $\rightarrow$ CVE-2026-27577}};
  \node[detail, draw=wmred, fill=wmredfill, text width=8.8cm,
        anchor=north, below=1.5mm of exploited.south] (high)
    {\textbf{HIGH RISK}\\[1pt]
     \texttt{RestElement}, \texttt{YieldExpression},
     \texttt{AwaitExpression},\\
     \texttt{ClassExpression}, \texttt{ImportExpression},
     \texttt{MetaProperty}};
  \node[detail, draw=wmorange, fill=wmorangefill, text width=8.8cm,
        anchor=north, below=1.5mm of high.south] (rest)
    {40 further unhandled types, each an unaudited instruction};

  \coordinate (botleft) at (\barsplit,0);
  \coordinate (botright) at (botleft |- rest.south);
  \draw[densely dashed, draw=wmgrey, line width=0.8pt]
    ($(botleft)+(0,-0.1)$) -- ($(botright)+(0,-0.15)$);
  \node[wmedgelabel, anchor=north, text=wmgrey,
        font=\sffamily\scriptsize\itshape]
    at ($(botright)+(0,-0.2)$) {enumeration boundary};

  \coordinate (midx) at (\barw/2,0);
  \coordinate (foot) at (midx |- rest.south);
  \node[wmnote, anchor=north, text width=13.0cm, below=9mm of foot]
    {Every unhandled node type is an instruction the sandbox's enumeration
     does not model. The specification grows; the enumeration does not.};

\end{tikzpicture}
\end{adjustbox}
\caption{AST coverage gap. The sandbox handles 25 of 72 ESTree node
types (35\%). The remaining 47 types constitute the weird machine's
instruction set, including the confirmed-exploitable SpreadElement.}
\label{fig:ast-coverage-gap}
\end{figure}

%% file: references.bib
@misc{ref1,
  author = {S. Bratus and M. E. Locasto and M. L. Patterson and L. Sassaman and A. Shubina},
  title = {{Exploit Programming: From Buffer Overflows to Weird Machines and Theory of Computation}},
  howpublished = {\emph{USENIX ;login:}, vol.~36, no. 6},
  year = {2011}
}

@misc{ref2,
  author = {S. Bratus and T. Darley and M. Locasto and M. L. Patterson and R. Shapiro and A. Shubina},
  title = {{Beyond Planted Bugs in `Trusting Trust': The Input-Processing Frontier}},
  howpublished = {\emph{IEEE Security \& Privacy}, vol.~12, no. 1, pp.~83-87},
  year = {2014}
}

@misc{ref3,
  author = {R. Shapiro and S. Bratus and S. W. Smith},
  title = {{\,`Weird Machines' in ELF: A Spotlight on the Underappreciated Metadata}},
  howpublished = {\emph{USENIX WOOT}},
  year = {2013}
}

@misc{ref4,
  author = {{GHSA-6cqr-8cfr-67f8}},
  title = {{Expression Escape Vulnerability Leading to RCE}},
  howpublished = {GitHub Security Advisory},
  year = {2026},
  url = {https://github.com/n8n-io/n8n/security/advisories/GHSA-6cqr-8cfr-67f8}
}

@misc{ref5,
  author = {{GHSA-75g8-rv7v-32f7}},
  title = {{Unauthenticated RCE via Form Node Double-Evaluation}},
  howpublished = {GitHub Security Advisory},
  year = {2026},
  url = {https://github.com/n8n-io/n8n/security/advisories/GHSA-75g8-rv7v-32f7}
}

@misc{ref6,
  author = {{GHSA-vpcf-gvg4-6qwr}},
  title = {{Expression Sandbox Escape Leading to RCE}},
  howpublished = {GitHub Security Advisory},
  year = {2026},
  url = {https://github.com/n8n-io/n8n/security/advisories/GHSA-vpcf-gvg4-6qwr}
}

@misc{ref7,
  author = {{GHSA-v98v-ff95-f3cp}},
  title = {{n8n Remote Code Execution via Expression Injection (CVE-2025-68613)}},
  howpublished = {GitHub Security Advisory},
  year = {2025},
  url = {https://github.com/n8n-io/n8n/security/advisories/GHSA-v98v-ff95-f3cp}
}

@misc{ref8,
  author = {{JEP 411}},
  title = {{Deprecate the Security Manager for Removal}},
  howpublished = {OpenJDK},
  year = {2021},
  url = {https://openjdk.org/jeps/411}
}

@misc{ref9,
  author = {{CVE-2026-22709}},
  title = {{vm2 sandbox escape}},
  howpublished = {NVD},
  year = {2026},
  url = {https://nvd.nist.gov/vuln/detail/CVE-2026-22709}
}

@misc{ref10,
  author = {A. Poutievski},
  title = {{Keeping your GitHub Actions and workflows secure: Preventing pwn requests}},
  howpublished = {GitHub Security Lab},
  year = {2021},
  url = {https://securitylab.github.com/resources/github-actions-preventing-pwn-requests/}
}

@misc{ref11,
  author = {E. Ilgayev},
  title = {{GitHub Actions exploitation: Repo jacking and environment manipulation}},
  howpublished = {Cycode},
  year = {2023},
  url = {https://cycode.com/blog/github-actions-vulnerabilities/}
}

@misc{ref12,
  author = {{ECMA-262}},
  title = {{ECMAScript Language Specification}},
  howpublished = {15th Edition, June},
  year = {2024},
  url = {https://tc39.es/ecma262/}
}

@misc{ref13,
  year = {n.d.},
  author = {{@n8n/tournament}},
  title = {{Expression evaluation engine for n8n}},
  howpublished = {GitHub.},
  url = {https://github.com/n8n-io/n8n/tree/master/packages/%40n8n/tournament}
}

@misc{ref14,
  author = {{Pillar Security}},
  title = {{n8n-poc: AI kill-chain proof of concept}},
  howpublished = {GitHub},
  year = {2026},
  url = {https://github.com/pillarsecurity/n8n-poc}
}

@misc{ref15,
  author = {{Pillar Security}},
  title = {{Critical Vulnerabilities in n8n Expose Hundreds of Thousands of Enterprise AI Systems to Complete Takeover}},
  howpublished = {Pillar Security Blog, February},
  year = {2026},
  url = {https://www.pillar.security/blog/n8n-sandbox-escape-critical-vulnerabilities-in-n8n-exposes-hundreds-of-thousands-of-enterprise-ai-systems-to-complete-takeover}
}

@misc{ref16,
  year = {n.d.},
  author = {{ESTree}},
  title = {{ESTree Spec}},
  howpublished = {GitHub.},
  url = {https://github.com/estree/estree}
}

@misc{ref17,
  year = {n.d.},
  author = {{ast-types}},
  title = {{Esprima-compatible implementation of the Mozilla JS Parser API}},
  howpublished = {GitHub.},
  url = {https://github.com/benjamn/ast-types}
}

@misc{ref18,
  author = {N. Nehorai},
  title = {{Achieving Remote Code Execution on n8n Via Sandbox Escape - CVE-2026-1470 \& CVE-2026-0863}},
  howpublished = {JFrog Security Research, January},
  year = {2026},
  url = {https://research.jfrog.com/post/achieving-remote-code-execution-on-n8n-via-sandbox-escape/}
}

@misc{ref19,
  author = {{CVE-2024-56326}},
  title = {{Jinja2 sandbox escape via indirect str.format reference}},
  howpublished = {NVD, December},
  year = {2024},
  url = {https://nvd.nist.gov/vuln/detail/CVE-2024-56326}
}

@misc{ref20,
  author = {{CVE-2025-27516}},
  title = {{Jinja2 sandbox escape via \textbar attr filter bypass}},
  howpublished = {NVD, March},
  year = {2025},
  url = {https://nvd.nist.gov/vuln/detail/CVE-2025-27516}
}

@misc{ref21,
  author = {{Retool}},
  title = {{Code Executor Sandbox Escape}},
  howpublished = {Retool Changelog Disclosures, July},
  year = {2025},
  url = {https://docs.retool.com/changelog/disclosures/sandbox-escape}
}

@misc{ref22,
  author = {L. Goderre},
  title = {{Security that strengthens the ecosystem: Docker's upstream approach to CVE-2025-12735}},
  howpublished = {Docker Blog},
  year = {2025},
  url = {https://www.docker.com/blog/security-that-strengthens-the-ecosystem-dockers-upstream-approach-to-cve-2025-12735/}
}

@misc{ref23,
  author = {M. Bargury},
  title = {{ZAPESCAPE: Organization-wide control over Code by Zapier}},
  howpublished = {Zenity, September},
  year = {2022},
  url = {https://zenity.io/blog/research/zapescape-organization-wide-control-over-code-by-zapier}
}
